\documentclass[hidelinks]{article}
\usepackage[utf8]{inputenc}
\usepackage[left=3cm,right=3cm,top=3cm,bottom=3cm]{geometry}
\usepackage{pdfpages}
\usepackage{hyperref}
\usepackage{listings}
\usepackage{xcolor}
\usepackage{setspace}
\usepackage{tocloft}
\usepackage{amsmath}
\usepackage{chngcntr}
\usepackage[toc,page]{appendix}
\usepackage[T1]{fontenc}
\usepackage[nottoc]{tocbibind}
\usepackage[compact]{titlesec}
\usepackage[utf8]{inputenc}
\usepackage{natbib}
\usepackage{amsmath}
\usepackage{amssymb}
\usepackage{amsfonts}

\usepackage{multirow}
\usepackage{setspace}
\usepackage{array}
\usepackage{float}

\newcolumntype{L}[1]{>{\raggedright\let\newline\\\arraybackslash\hspace{0pt}}m{#1}}
\newcolumntype{C}[1]{>{\centering\let\newline\\\arraybackslash\hspace{0pt}}m{#1}}
\newcolumntype{R}[1]{>{\raggedleft\let\newline\\\arraybackslash\hspace{0pt}}m{#1}}

\usepackage{subfig}
\usepackage{bm}
\usepackage{multirow}
\usepackage{tikz}
\usetikzlibrary{shapes,arrows}
\usepackage{blkarray}
\usepackage[affil-it]{authblk}
\usepackage{listings}
\usepackage{siunitx}
\usepackage{lineno}

\usepackage{natbib}
\usepackage{color}
\providecommand{\keywords}[1]{\textbf{\textit{Keywords ---}} #1}

\title{A doubly self-exciting Poisson model for describing scoring levels in NBA basketball}
\author[1*]{Álvaro Briz-Redón}
\date{}
\affil[1]{Department of Statistics and Operations Research, University of Valencia, Spain}
\affil[*]{\textsf{alvaro.briz@uv.es}}

\begin{document}

\maketitle







\begin{abstract}

In this paper, Poisson time series models are considered to describe the number of field goals made by a basketball team or player at both the game (within-season) and the minute (within-game) level. To deal with the existence of temporal autocorrelation in the data, the model is endowed with a doubly self-exciting structure, following the INGARCH(1,1) specification. To estimate the model at the within-game level, a divide-and-conquer procedure, under a Bayesian framework, is carried out. The model is tested with a selection of NBA teams and players from the 2018-2019 season.

\end{abstract}

\keywords{Bayesian inference, NBA data, Poisson model, self-exciting model, sports analytics}

\section{Introduction}

The statistical analysis of datasets arising from the practice of professional sports has gained interest in recent years, both from a methodological and practical point of view. As a consequence, the number of publications on this subject in statistical journals shows an increasing trend \citep{swartz2020should}. Specifically, basketball datasets are receiving great attention from researchers. Among the topics covered, we can highlight the pre-game and in-game prediction of game outcomes \citep{shi2021discrete,song2020modelling,tian2020modeling}, the estimation of the spatial distribution of shot locations \citep{yin2020bayesian}, the clustering of shot selection spatial structures \citep{hu2021bayesian,yin2022analysis}, the study of performance variability \citep{sandri2020markov}, the analysis of tracking data \citep{santos2022role}, and the construction of network competition models \citep{horrace2022network}.

Another topic of great relevance in the context of basketball analytics is the one known as the ``hot hand'' hypothesis, which gained scientific attention after the study of \cite{gilovich1985hot}. The hot hand refers to the increase in a player's probability of success after one or several successful shots. This hypothesis has been studied more extensively with free throws and to a lesser extent in the case of field goals. In a recent paper, \cite{lantis2021hot} have found a small hot hand effect for free throws and no effect for field goals, even though the existing literature presents mixed results on this subject.

The main variable for the present study has been the number of field goals made by a team or player over a time partition, which leads us to the analysis of time series data. Since the number of field goals made is an integer-valued variable, the Poisson model is a suitable choice for the analysis. Specifically, in this paper, we present a Poisson model with a self-exciting structure that allows us to analyze the temporal dependence in scoring levels shown by a basketball team or player. Self-exciting models, which are widely used and applied in some fields such as criminology \citep{mohler2011self} and earthquake analysis \citep{ogata1986point}, allow accounting for the impact of the history of the process on the variable of interest, that is, the effect of (recent) past observations on its expected value at later time points. In the present paper, we consider a Poisson model with an autoregressive conditional heteroskedastic structure, which has been typically used for the analysis of economic time series \citep{comte2000second}, although it has also started to be adopted in other contexts, such as criminology \citep{clark2021class,escudero2022spatially} and sports analytics \citep{cerqueti2022ingarch}. More precisely, a doubly self-exciting specification is chosen to analyze the data at two different temporal resolution levels: at the game level (within a season) and the minute level (within a game).

The paper is structured as follows. Section \ref{Data} describes the datasets used for the analysis. Section \ref{Methodology} describes the modeling approach proposed, and how model estimation and evaluation are carried out. A secondary use of the outcomes of this model for time series clustering is also depicted. Then, Section \ref{Results} includes the main results obtained for the datasets considered. Finally, Section \ref{Discussion} provides some concluding remarks.

\section{Data}
\label{Data}

A dataset containing information about all field goals made and missed during the 2018-19 NBA season has been used for the analysis. This dataset was downloaded from \url{https://datavizardry.com/2020/01/28/nba-shot-charts-part-1/}. This kind of dataset can be constructed with the aid of the NBA application programming interface (API). Specifically, for each made/missed field goal during the season, the dataset includes, among other variables, the name of the player who shot, the player's team, and the quarter and minute (within the quarter) when the shot took place. For the analysis presented in the present paper, only made field goals are considered. Besides, the temporal location of each made shot is transformed into the minute of the game within which the made field goal occurred. Thus, an integer number between 1 and 48 is assigned to each element of the dataset. Field goals made on overtime have been discarded for analysis.

\begin{figure}[hbt]
 \centering
 \subfloat[]{\includegraphics[width=5cm,angle=-90]{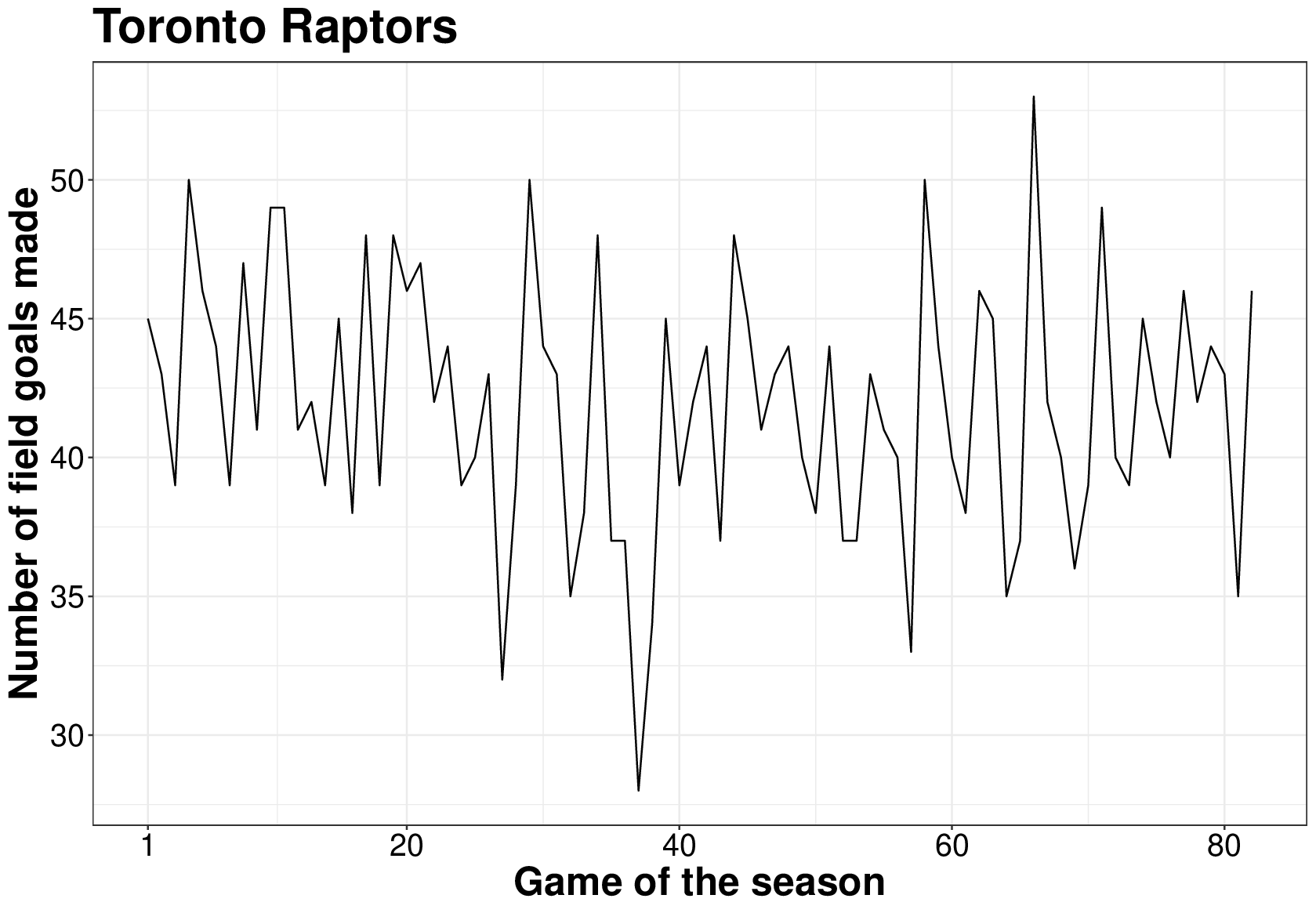}\label{fig:data_description_a}}
 \subfloat[]{\includegraphics[width=5cm,angle=-90]{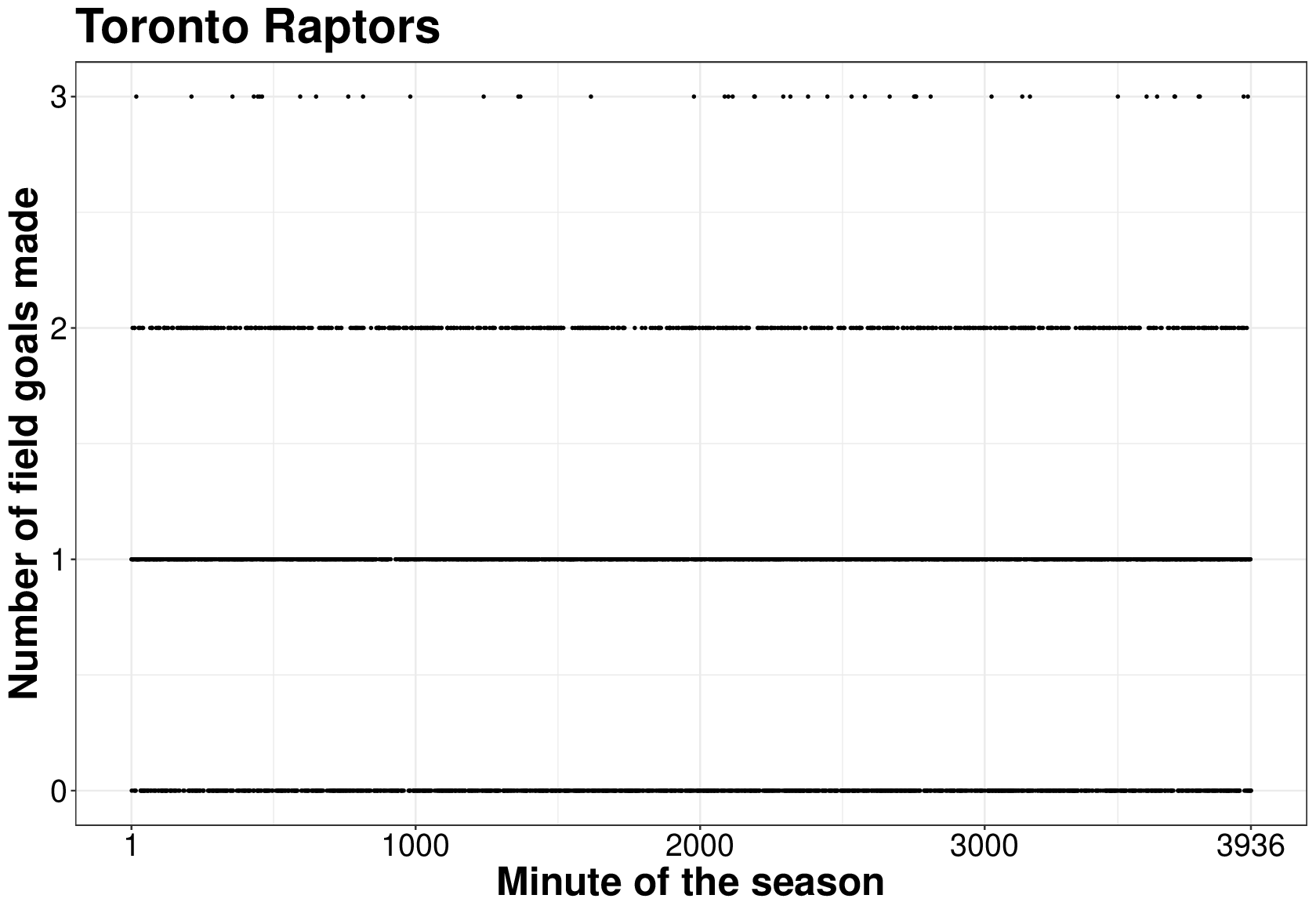}\label{fig:data_description_b}}\\
  \subfloat[]{\includegraphics[width=5cm,angle=-90]{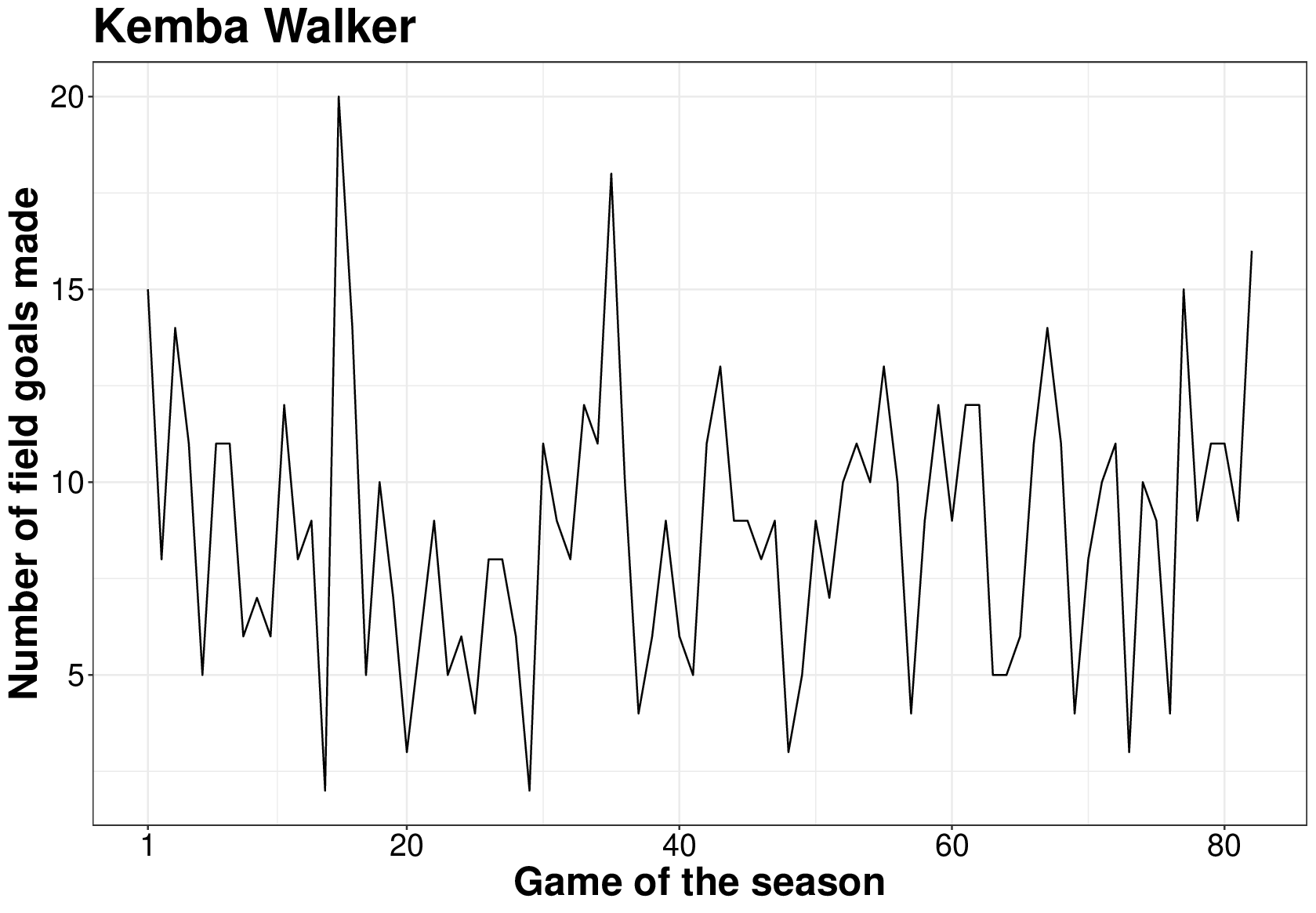}\label{fig:data_description_c}}
 \subfloat[]{\includegraphics[width=5cm,angle=-90]{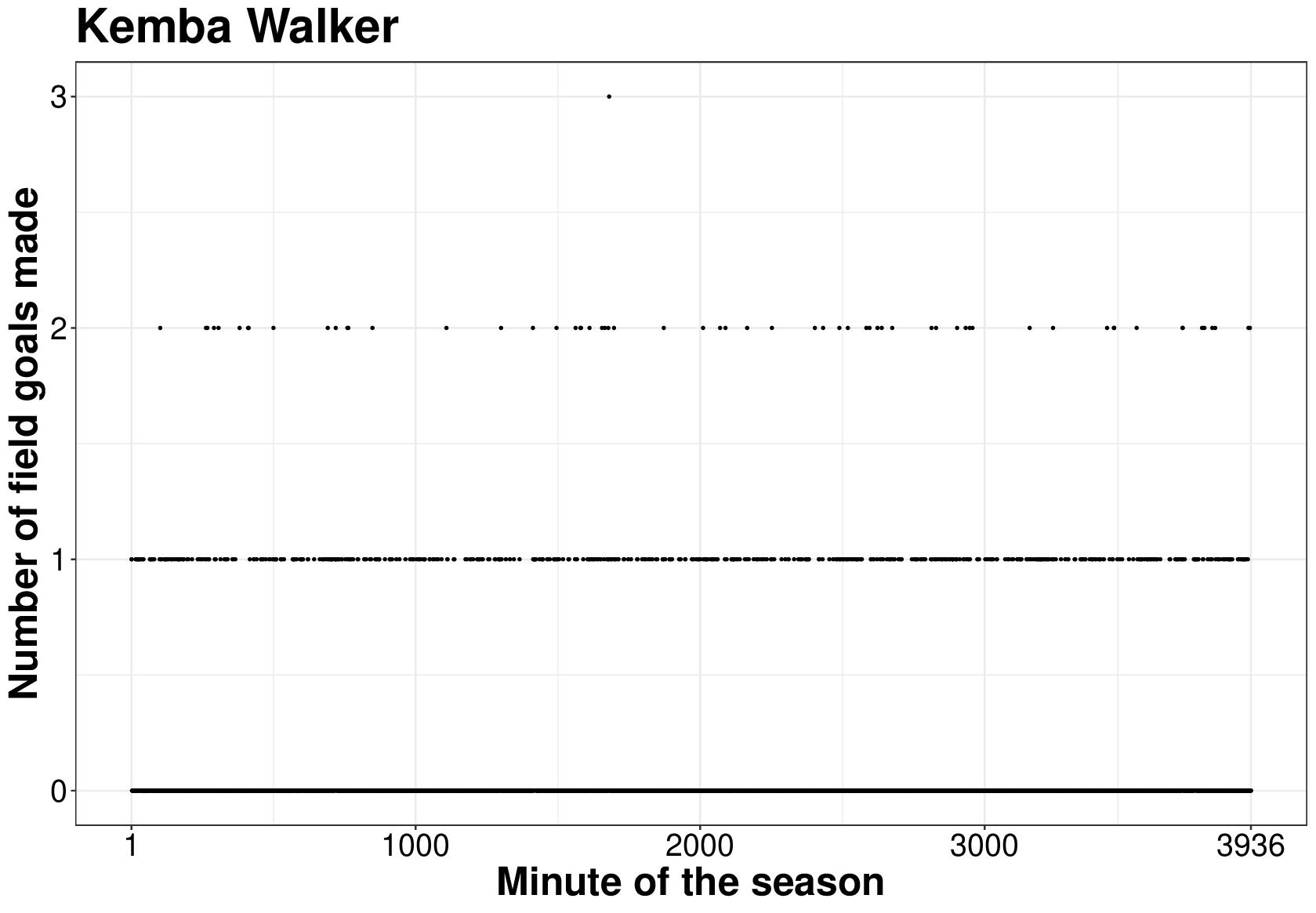}\label{fig:data_description_d}}\\
\caption{Graphical description of the response variables considered for the analysis: number of field goals made per game ((a) and (c)), and number of field goals made per minute ((b) and (d)). The data correspond to the 2018-19 season of the Toronto Raptors ((a) and (b)) and Kemba Walker ((c) and (d))}
\label{fig:data_description}
\end{figure}

A total of 30 teams compete in the current NBA league, which is divided into two conferences, the Eastern and the Western. The competition system consists of 82 regular season games and a play-off in which the eight best-ranked teams from each conference at the end of the regular season participate. For this study, the eight teams that reached their respective Conference Semifinals in the 2018-19 season were considered: Boston Celtics (BOS), Denver Nuggets (DEN), Golden State Warriors (GSW), Houston Rockets (HOU), Milwaukee Bucks (MIL), Philadelphia 76ers (PHI), Portland Trail Blazers (POR), and Toronto Raptors (TOR). In addition, we have analyzed the scoring levels of eight players of that season: Bradley Beal (BB), Damian Lillard (DL), Donovan Mitchell (DM), James Harden (JH), Karl-Anthony Towns (KAT), Kemba Walker (KW), Kevin Durant (KD), and Paul George (PG). These players correspond to the top scorers of the 2018-19 regular season, considering the NBA players who participated in more than 75 games of that season. In the following, $N_g$ will represent the number of games played by a team or player during that season. Therefore, it holds that $N_g = 82$ for all the teams and $N_g \in \{76,..., 82\}$ for each of the players. Figure \ref{fig:data_description} shows the scoring pattern, in terms of the number of field goals made per game and minute of the season, for the Toronto Raptors and Kemba Walker.
                     
\section{Methodology}
\label{Methodology}

The methodological objective of the present work is to define a model capable of capturing the time dependence in our variable of interest, the number of field goals made by a team or player, and the self-exciting nature that the stochastic process driving this variable might present. Besides, we are interested in modeling the number of field goals made over two different temporal schemes: the games within an NBA season, and the minutes within an NBA game. Therefore, hereinafter we will talk about the within-season and within-game variation of the data accordingly.

\subsection{The INGARCH model}

The INGARCH model \citep{ferland2006integer} is a time series model that corresponds to the integer-valued generalized autoregressive conditional heteroskedastic (GARCH) process, which was previously introduced by \cite{bollerslev1986generalized}. If $\{Y_t\}_{t\in\mathcal{T}}$ represents a time series of counts indexed over a time partition $\mathcal{T}$, we can assume $Y_t$ follows a Poisson distribution with a rate $\lambda_t$ and model this rate. The structure of $\lambda_t$ under the INGARCH($p$,$q$) model is
\begin{align*}
Y_t  &\sim Po(\lambda_t) \\
\lambda_t &= d + \sum_{i=1}^{p} \kappa_{i} \lambda_{t-i} + \sum_{j=1}^{q} \eta_{j} Y_{t-j}\text{.}
\end{align*}

This type of model weights the effect of past observations with that of the expected value of the time series, which can depend on exogenous covariate information. The $\eta_i$'s represent repetition in the stochastic process, whereas the $\kappa_j$'s control the decay in $Y_t$ under the absence of repetition. A particular case of this model, which is widely used for practical reasons, is the INGARCH(1,1) specification, which corresponds to the expression
\begin{align*}
Y_t  &\sim Po(\lambda_t) \\
\lambda_t &= d + \kappa \lambda_{t-1} + \eta Y_{t-1}\text{.}
\end{align*}

The main properties of this model were already established by \cite{heinen2003modelling}. Provided that $\kappa+\eta<1$, the INGARCH(1,1) model is stationary and its unconditional mean is
$$E(Y_{t})=\mu=\frac{d}{1-(\kappa+\eta)}\text{,}$$
whereas the unconditional variance of $Y_t$ is
$$Var(Y_{t})=\frac{\mu(1-(\kappa+\eta)^2+\eta^2)}{1-(\kappa+\eta)^2}\text{.}$$

Therefore, it holds that $Var(Y_{t}) \geq E(Y_{t})$, which tells us that the model allows capturing the overdispersion of the observations (as long as $\eta\neq0$), despite its (conditional) definition as a Poisson model. Particularly, the overdispersion of the model is an increasing function of $\eta$ and, to a lesser extent, of $\kappa$. Finally, the unconditional autocorrelation function is given by
\begin{equation} \label{eq:autocorrelation_function}
Corr(Y_{t},Y_{t-r})=\frac{\eta(1-\kappa(\kappa+\eta))(\kappa+\eta)^{r-1}}{1-(\kappa+\eta)^2+\eta^2}\text{.}
\end{equation}

\subsection{A doubly self-exciting Poisson model}

\subsubsection{Game-level model}

A doubly self-exciting Poisson model is proposed to describe, simultaneously, the number of field goals made by a basketball team/player over the games within an NBA season, and the minutes within each game of the season. Let $\{Y_g\}_{g\in\{1, ..., N_g\}}$ be the time series representing the number of field goals made by a basketball team/player over the $N_g$ games played by a team/player during an NBA season, and let $\{Y_{gm}\}_{g\in\{1, ..., N_g\},m\in\{1, ..., 48\}}$ be the time series representing the number of field goals made by a basketball team/player over the ($N_g\cdot48$) minutes of corresponding to the $N_g$ games played by the team or player (in the case of teams, since $N_g=82$, the length of the time series is $82\cdot48=3936$). 

If we assume that $Y_g \sim Po(\lambda_g)$, where $\lambda_g$ represents the scoring level at game $g$, the INGARCH model assumes that the expected number of field goals made by a basketball team/player in a game depends on previously expected values (the $\lambda_g$'s) and previously observed values (the $Y_g$'s). This allows accounting for the self-exciting nature of the data under analysis. Besides, the structure of the INGARCH model admits the inclusion of fixed and random effects. In the present paper, the following INGARCH(1,1) structure is considered for modeling the $Y_g$'s:
\begin{equation} \label{eq:double_self_S}
\begin{split}
Y_g  &\sim Po(\lambda_g) \\
\lambda_g &= \exp(\alpha_S+\alpha_{\text{Home}}I_{Home_{g}=1})+\kappa_S \lambda_{g-1}I_{g>1}+\eta_S Y_{g-1} I_{g>1}\text{,}
\end{split}
\end{equation}
where $\alpha_S$ is a season-level intercept parameter, $\alpha_{\text{Home}}$ represents the effect of playing at home on the response variable, $\kappa_S$ represents the influence of the previous expectation, $\lambda_{g-1}$, on $\lambda_g$, and $\eta_S$ the influence of the previous observation, $Y_{g-1}$, on $\lambda_g$. The term $I_{g>1}$ represents an indicator function that takes value 1 for $g>1$. For $g=1$ we cannot include the self-exciting parameters in the specification of $\lambda_1$ (there is no observed history of the process for $g=1$), so only the $\alpha_S$ and $\alpha_{\text{Home}}$ parameters participate in the estimation.


\subsubsection{Minute-level model}

In a second step, an analogous INGARCH structure is considered for describing the expected number of field goals made by a basketball team/player per minute over the whole NBA season. In addition, the term $\lambda_g$ modeled according to (\ref{eq:double_self_S}) is included in the specification of $\lambda_{gm}$, which enables us to consider the (self-exciting) scoring level of a team/player on game $g$. Specifically, the following expression is employed for modeling $\lambda_{gm}$
\begin{equation} \label{eq:double_self_G}
\begin{split}
Y_{gm} &\sim Po(\lambda_{gm}) \\
\lambda_{gm} &= \exp\biggl(\alpha_G+\sum_{Q\in \{1,2,3,4\}} \sum_{H\in H_{Q}} \alpha_{QH}I_{QH_{gm}=1}+\frac{\lambda_g}{48}\biggl)+\kappa_G \lambda_{gm-1} I_{m>1}+\eta_G Y_{gm-1}I_{m>1}\text{.}
\end{split}
\end{equation}
Parameters $\alpha_G$, $\kappa_G$, and $\eta_G$ represent for (\ref{eq:double_self_G}), at the game level, the same than $\alpha_S$, $\kappa_S$, and $\eta_S$ do in (\ref{eq:double_self_S}). It is worth noting that the modeled expected value on game $g$, $\lambda_g$, is divided by 48 to be included in the specification of $\lambda_{gm}$. This is a necessary correction since $\lambda_g$ represents an expected scoring level per game, not per minute. The term $I_{m>1}$ represents an indicator function that takes value 1 for $m>1$. In a similar way to what happens to the game-level model, we cannot include the self-exciting parameters in the specification of the $\lambda_{g1}$'s, leaving $\alpha_G$ and the $\alpha_{QH}$'s as the only parameters involved in their estimation.

Therefore, this modeling approach presents a hierarchical structure, in the sense that the expected scoring level at a minute $m$ of game $g$ depends on the (overall) expected scoring level at game $g$. As both $\lambda_{g}$ and $\lambda_{gm}$ do follow a self-exciting specification, the full model is referred to in the remainder of the paper as a doubly self-exciting Poisson model. Strictly speaking, the minute-level model would be the one endowed with a doubly self-exciting structure, whereas the game-level model should be better described as a self-exciting (INGARCH(1,1)) model. Nevertheless, the term doubly self-exciting Poisson model would be used indistinctly throughout the paper, as the outputs from the game-level model are also implicated in the estimation of the minute-level modeling process. 

\subsection{Model estimation}

Both the game-level and the minute-level models are fitted within a Bayesian framework. Thus, uninformative $N(0,1000)$ priors have been placed on the intercept parameters ($\alpha_S$ and $\alpha_G$) and the fixed-effects parameters ($\alpha_{\text{Home}}$ and the $\alpha_{QH}$'s). Besides, a uniform $U(0,1)$ prior has been assumed for the parameters representing self-exciting effects ($\eta_{S}$, $\kappa_{S}$, $\eta_{G}$, and $\kappa_{G}$). The models have been fitted with the aid of the NIMBLE system for Bayesian inference \citep{de2017programming}, which is based on Monte Carlo Markov Chain (MCMC) procedures. 

At this point, it is important to remark that fitting the model proposed is computationally demanding, so some strategies have been carried out for model estimation. First, the game-level and the minute-level model are estimated separately. We start by estimating the game-level model, which allows us to compute the posterior mean of $\lambda_g$, for $g=1,...,N_g$. The posterior mean of $\lambda_g$, which we denote by $\hat{\lambda}_g$, is then considered as an offset term in the minute-level model:
\begin{equation} \label{eq:double_self_G_est}
\begin{split}
Y_{gm} &\sim Po(\lambda_{gm}) \\
\lambda_{gm} &= \exp\biggl(\alpha_G+\sum_{Q\in \{1,2,3,4\}} \sum_{H\in H_{Q}} \alpha_{QH}I_{QH_{gm}=1}+\frac{\hat{\lambda}_g}{48}\biggl)+\kappa_G \lambda_{gm-1} I_{m>1}+\eta_G Y_{gm-1}I_{m>1}\text{,}
\end{split}
\end{equation}
which is estimated subsequently. In order to fit this model, which is the one that presents the main challenge at a computational level, a divide-and-conquer strategy is followed. We consider the following partition of the observations:
$$\{Y_{gm}\}_{g\in G}=\bigcup_{k=1}^{K} \{Y_{gm}\}_{g\in G_{k}}\text{,}$$
where $G$ denotes the set of all the games played by the NBA team/player, and $\{G_{k}\}_{k=1}^{K}$ represents a non-overlapping partition of $G$. In all cases, it holds $m\in\{1, ..., 48\}$, so data values from a specific game are not shared across partitions. In order to merge the posterior distributions of a parameter of interest corresponding to each subset $G_k$, the Wasserstein barycenter \citep{cuturi2014fast} of the set formed by all the posterior distributions derived from MCMC simulations is computed. This method has good theoretical properties, as shown by \cite{ou2021scalable}. Hence, if $p(\theta|\{Y_{gm}\}_{g\in G_{k}})$ represents the posterior distribution of a parameter of the model ($\theta$) based on subset $G_k$, the combined season-level posterior distribution of $\theta$ is computed as
$$\bar{p}(\theta|\{Y_{gm}\}_{g\in G})=WB(p(\theta|\{Y_{gm}\}_{g\in G_{1}}),...,p(\theta|\{Y_{gm}\}_{g\in G_{K}}))\text{,}$$
where $WB$ denotes the Wasserstein barycenter. In the remainder of the paper, we will refer to $\bar{p}(\theta|\{Y_{gm}\}_{g\in G})$ as the barycentric posterior distribution of the corresponding parameter.

\subsection{Model comparison}

Model comparison has been performed through the Watanabe–Akaike Information Criterion (WAIC) proposed by \cite{watanabe2010asymptotic}. The WAIC is a measure of the goodness-of-fit of a Bayesian model while accounting for its complexity in terms of the number of effective parameters involved in the model. The WAIC allows for direct model comparison, given a set of models fitted to the same data. As a general rule, the model with the smallest WAIC value is the one that shows the greatest performance, meaning the best balance between fit and complexity. Each of the proposed models, at both the game and the minute level, is compared against a baseline model that excludes self-exciting effects. The structure of the baseline models used in each case is outlined in the Results section.

\subsection{Clustering scoring levels}

As a secondary analysis, we propose to use the outputs of the doubly self-exciting Poisson model to perform a hierarchical clustering of the analyzed teams or players. Recently, \cite{cerqueti2022ingarch} have proposed a new fuzzy clustering method for time series based on the fit of INGARCH(1,1) models. In our case, we propose to use the Wasserstein distance employed for the calculation of the barycentric posterior distribution to measure the similarity between teams or players, in terms of some model parameter. In particular, in our case, the interest lies especially in the parameters involved in the self-exciting component of the model. This allows computing a dissimilarity matrix for the set of teams or players, and performing a hierarchical clustering of these elements following the common methods available in this regard. Specifically, we consider the unweighted pair group method with arithmetic mean (UPGMA) introduced by \cite{sokal1958statistical}, which is implemented in the $\textsf{hclust}$ R function.

\subsection{Software}

The R programming language \citep{teamR} has been used for the analysis. In particular, the R packages \textsf{ggplot2} \citep{ggplot2}, \textsf{nimble} \citep{de2017programming}, \textsf{transport} \citep{transport}, and \textsf{T4transport} \citep{T4transport} have been used.










\section{Results}
\label{Results}

This section discusses the main results derived from the analysis of the teams and players selected with on the modeling framework proposed. First, we analyze the estimates obtained for the fixed effects of both models, i.e., the effect of playing at home for the game-level model, and the temporal effect given by the half-quarters in the case of the minute-level model. Second, we analyze the self-exciting effects included in the model, both at the game and the minute level, which lies the main motivation of the study.

\subsection{Playing-at-home effect}

\begin{figure}[hbt]
 \centering
 \subfloat[]{\includegraphics[width=5cm,angle=-90]{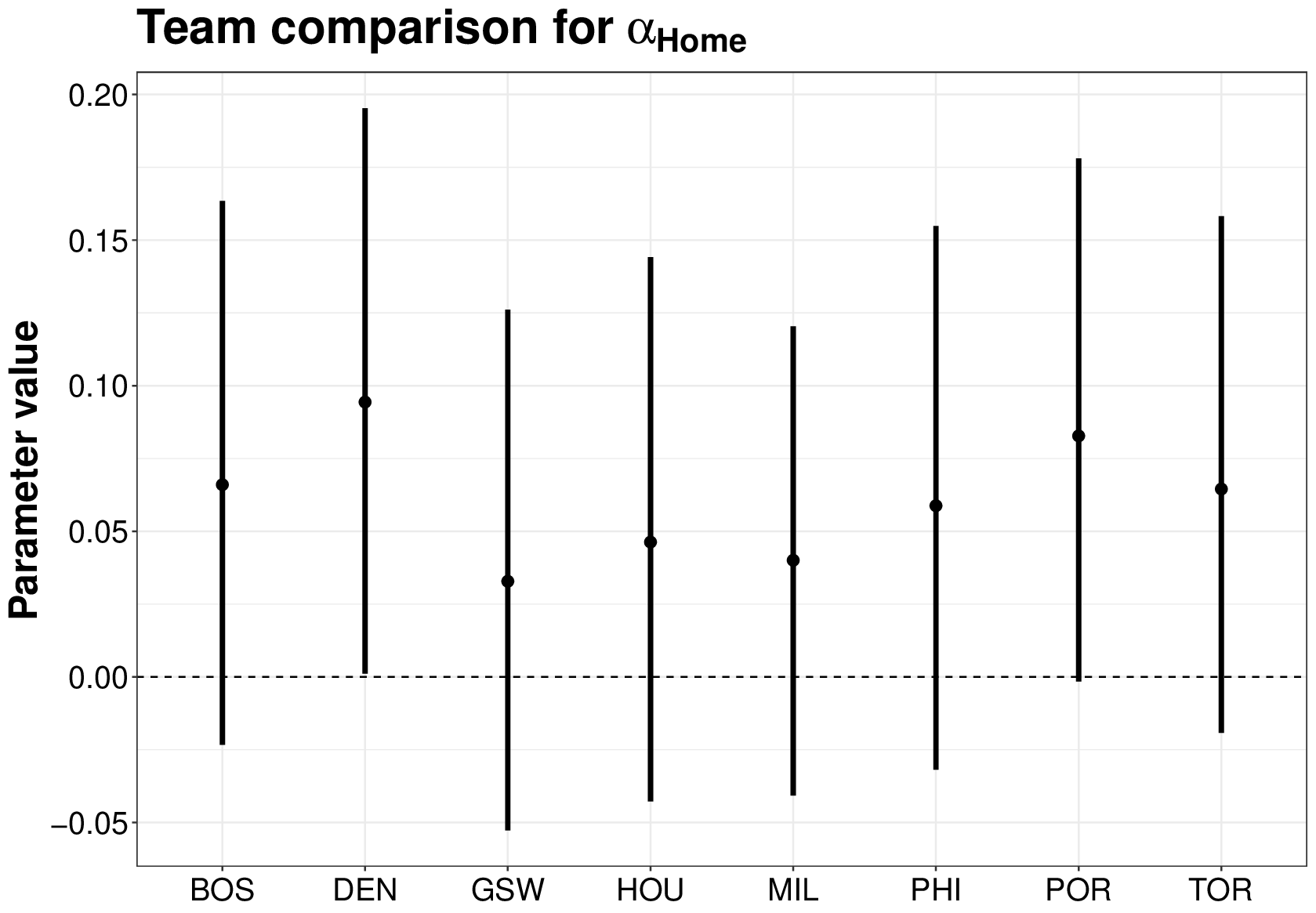}\label{fig:alpha_home_teams_a}}
 \subfloat[]{\includegraphics[width=5cm,angle=-90]{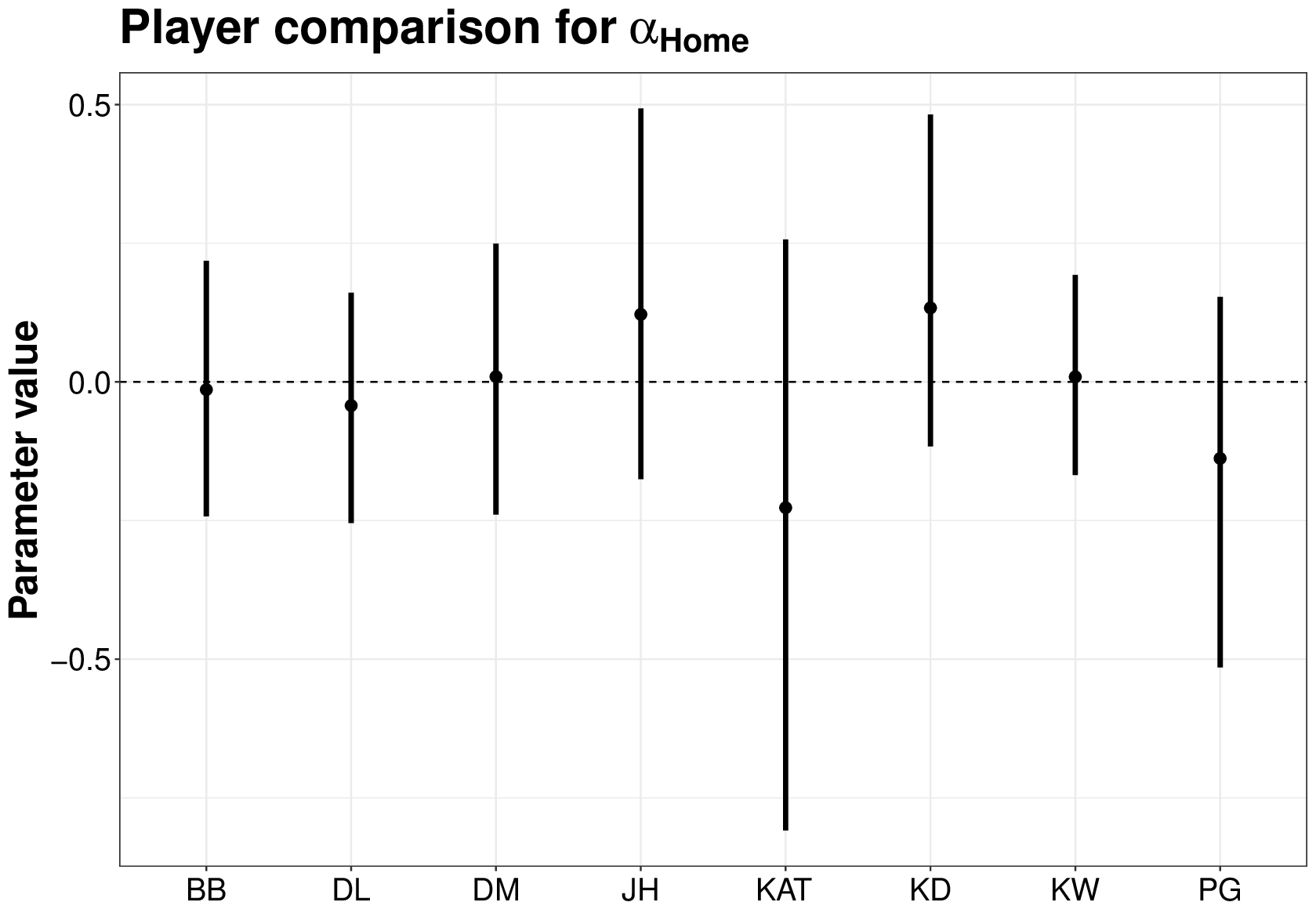}\label{fig:alpha_home_teams_b}}
\caption{Estimate of $\alpha_{\text{Home}}$, computed as the mean value of the posterior distribution of the parameter, and associated 95\% credible intervals for the teams (a) and players (b) under study}
\label{fig:alpha_home_teams}
\end{figure}

Whether or not there is an advantage for the home team, in terms of the probability of winning a game, is a widely studied issue. In general, previous studies suggest that this effect exists, although it depends on the team \citep{pollard2007home} and the professional league (country) under analysis \citep{gomez2011reduced}. This paper analyzes if the scoring level depends on whether the team plays at home or not. This point is also studied for the players considered for analysis since the variable indicating whether the match is played at home or away is included in the model for players as well. Thus, Figure \ref{fig:alpha_home_teams} summarizes the estimates obtained for the effect of playing at home (measured through parameter $\alpha_{\text{Home}}$) on the number of field goals made per game, both at the team (Figure \ref{fig:alpha_home_teams_a}) and player level (Figure \ref{fig:alpha_home_teams_b}). The results at the team level suggest that playing at home might increase the number of field goals made per game since the posterior mean of $\alpha_{\text{Home}}$ is greater than 0 in all cases. However, only DEN and POR show a clear effect, considering the associated 95\% credible intervals, as shown in Figure \ref{fig:alpha_home_teams_a}. In the case of DEN, this interval does not contain 0, whereas for POR the lower bound of this interval is slightly below 0. In the case of the players analyzed, no playing-at-home advantage is observed for any of them, as can be seen in Figure \ref{fig:alpha_home_teams_b}. Only two of the players studied, JH and KD, present a posterior mean above 0, while KAT and PG display the opposite behavior, even though these results cannot be regarded as statistically relevant. Finally, the posterior distribution of the remaining four players that have been analyzed is situated around 0, suggesting the absence of an advantage (or disadvantage) from playing at home in terms of scoring levels.

\subsection{Within-game temporal effects}

\begin{figure}[hbt]
 \centering
 \subfloat[]{\includegraphics[width=5cm,angle=-90]{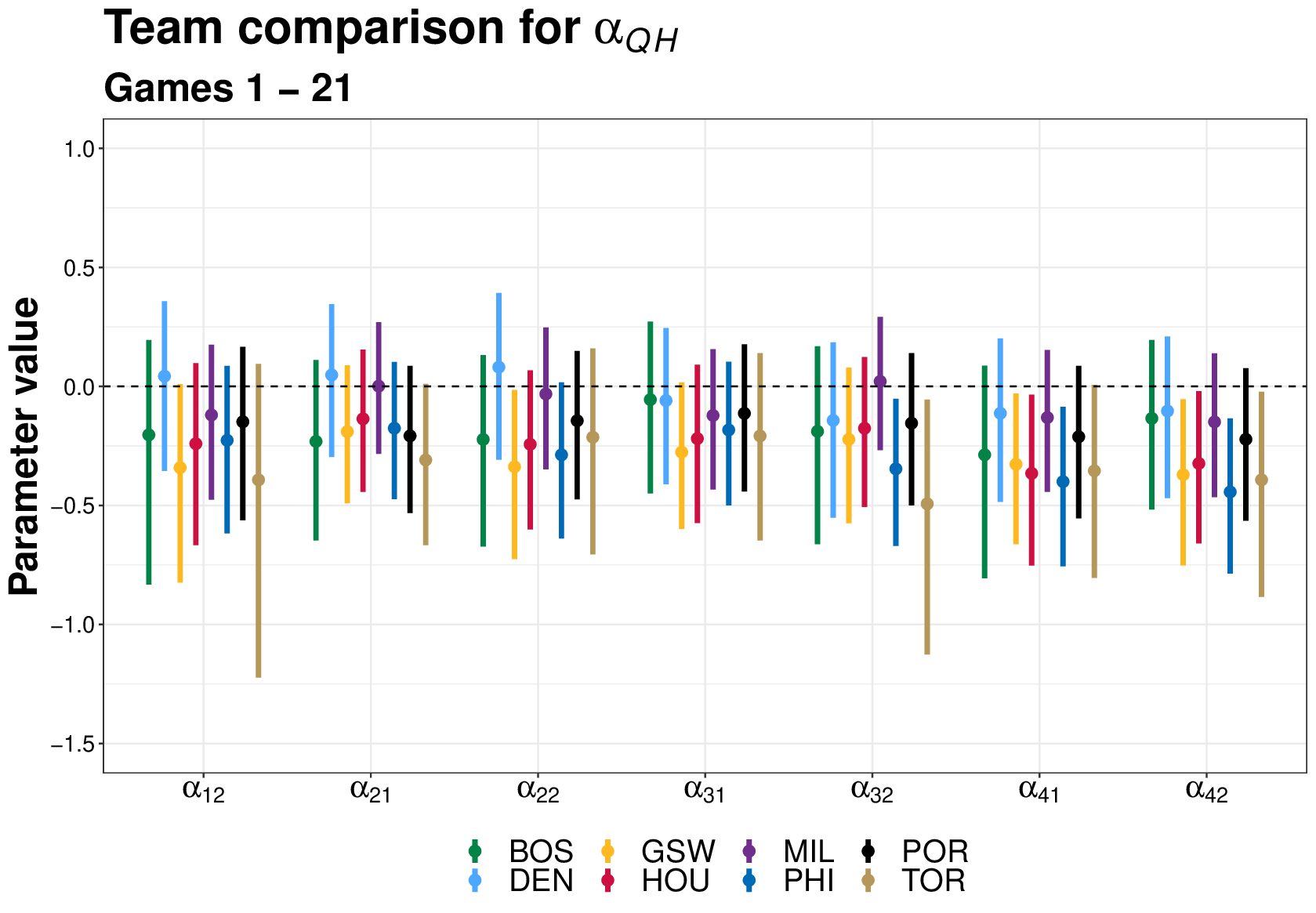}\label{fig:Quarter_half_team_comparison_a}}
 \subfloat[]{\includegraphics[width=5cm,angle=-90]{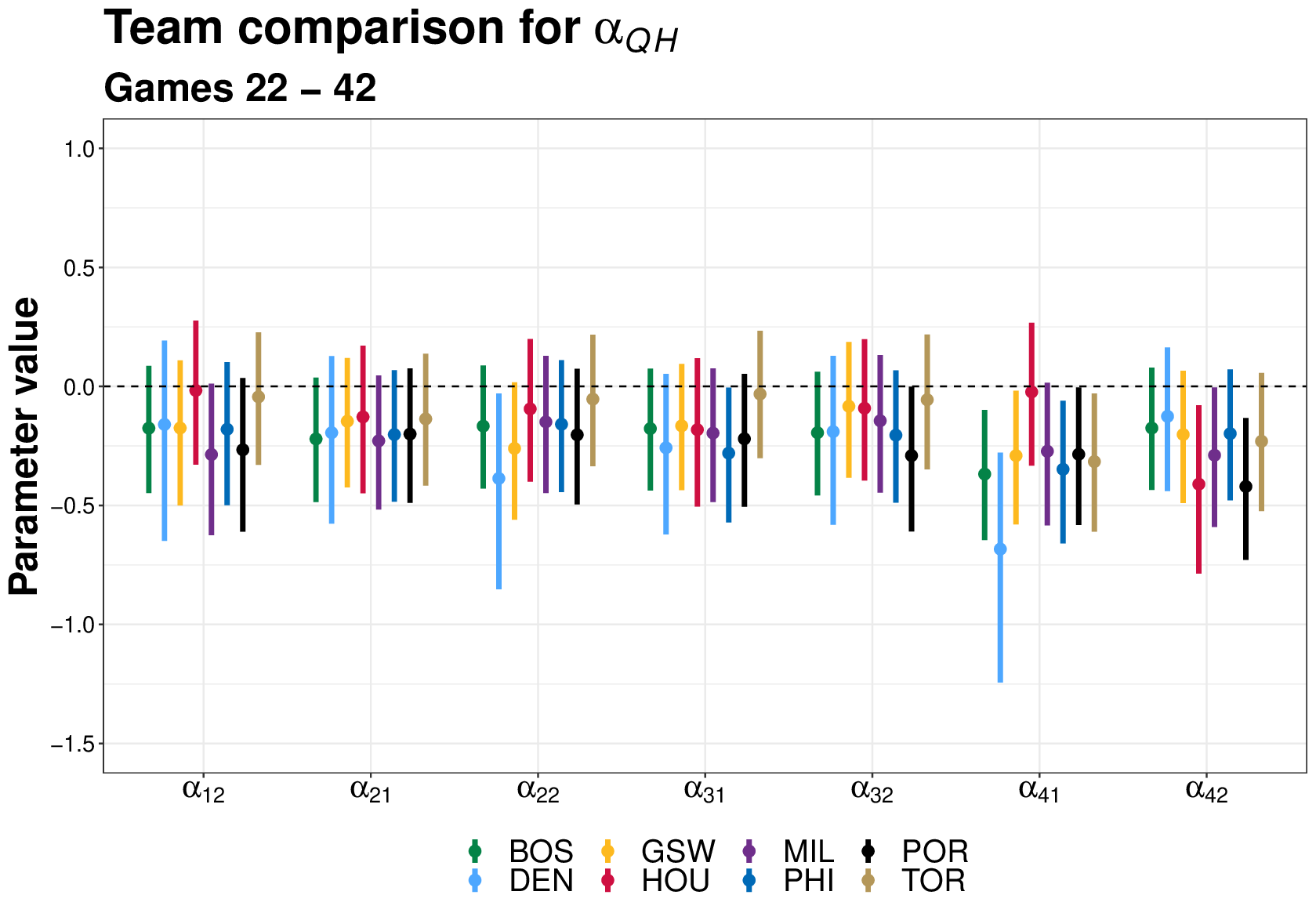}\label{fig:Quarter_half_team_comparison_b}}\\
  \subfloat[]{\includegraphics[width=5cm,angle=-90]{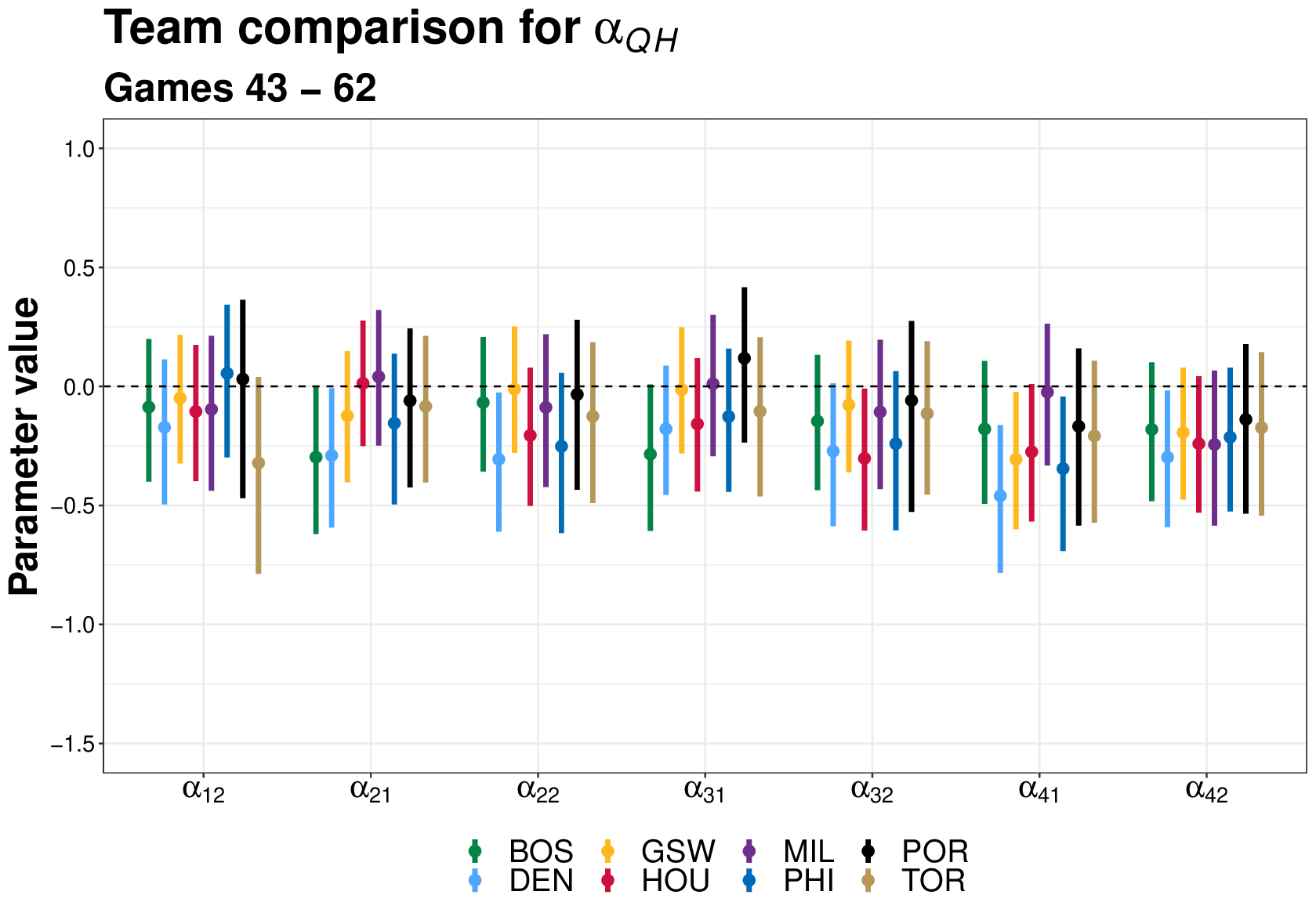}\label{fig:Quarter_half_team_comparison_c}}
 \subfloat[]{\includegraphics[width=5cm,angle=-90]{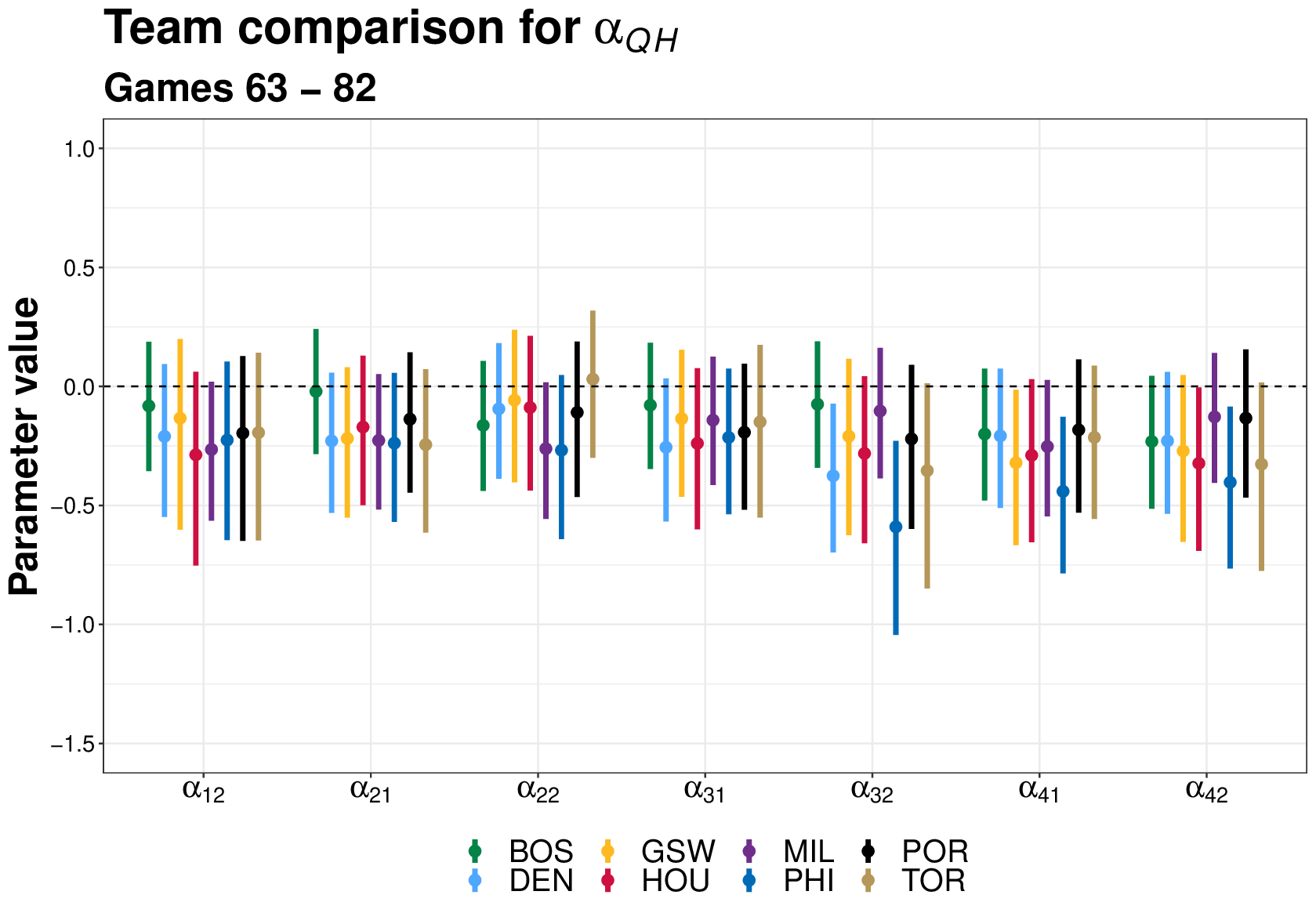}\label{fig:Quarter_half_team_comparison_d}}
\caption{Estimates of the $\alpha_{QH}$'s, computed as the mean value of the posterior distribution of the parameter, and associated 95\% credible intervals for the teams under study. The estimates are provided for each of the periods of the season considered for model inference}
\label{fig:Quarter_half_team_comparison}
\end{figure}

\begin{figure}[hbt]
 \centering
 \subfloat[]{\includegraphics[width=5cm,angle=-90]{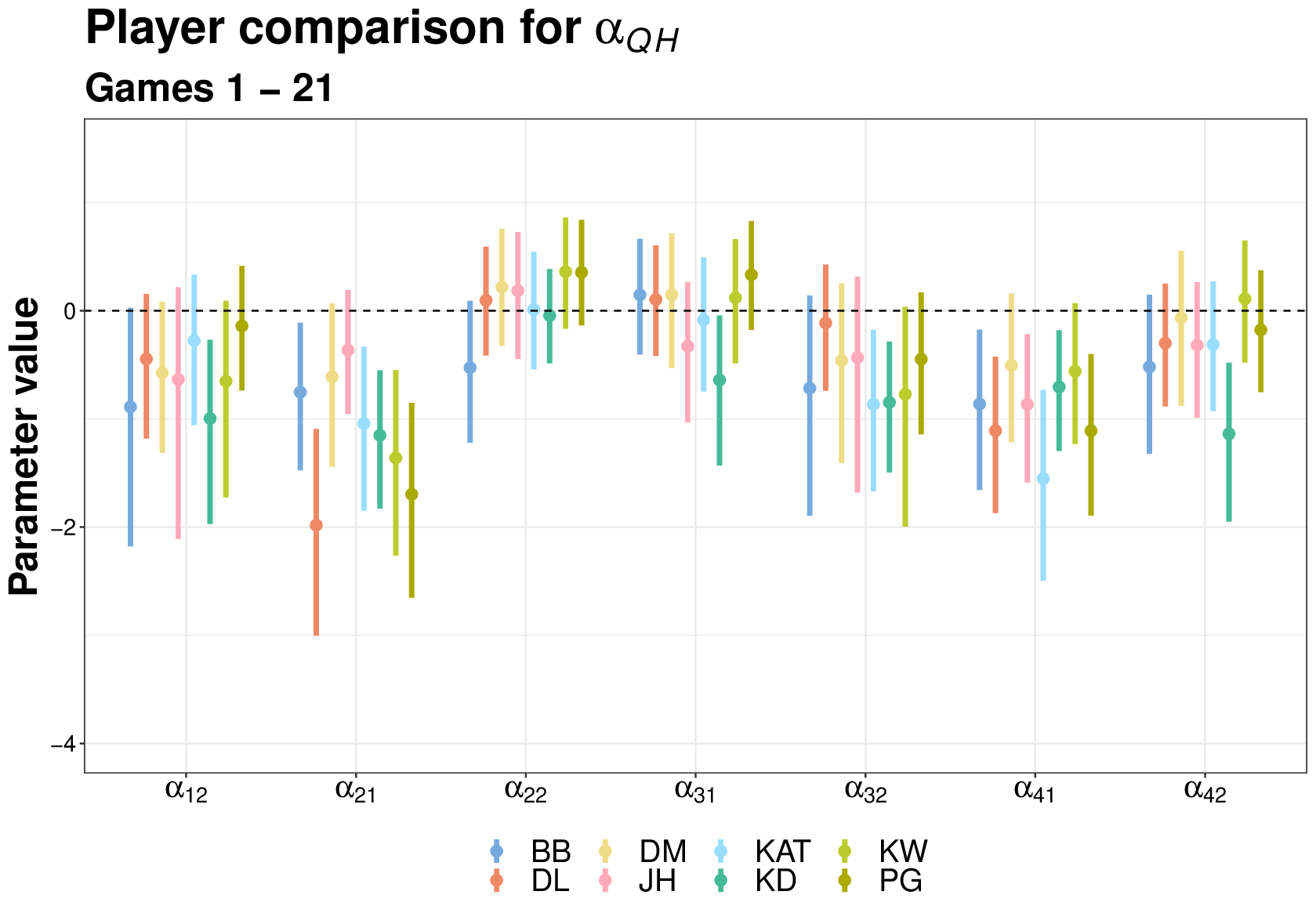}\label{fig:Quarter_half_player_comparison_a}}
 \subfloat[]{\includegraphics[width=5cm,angle=-90]{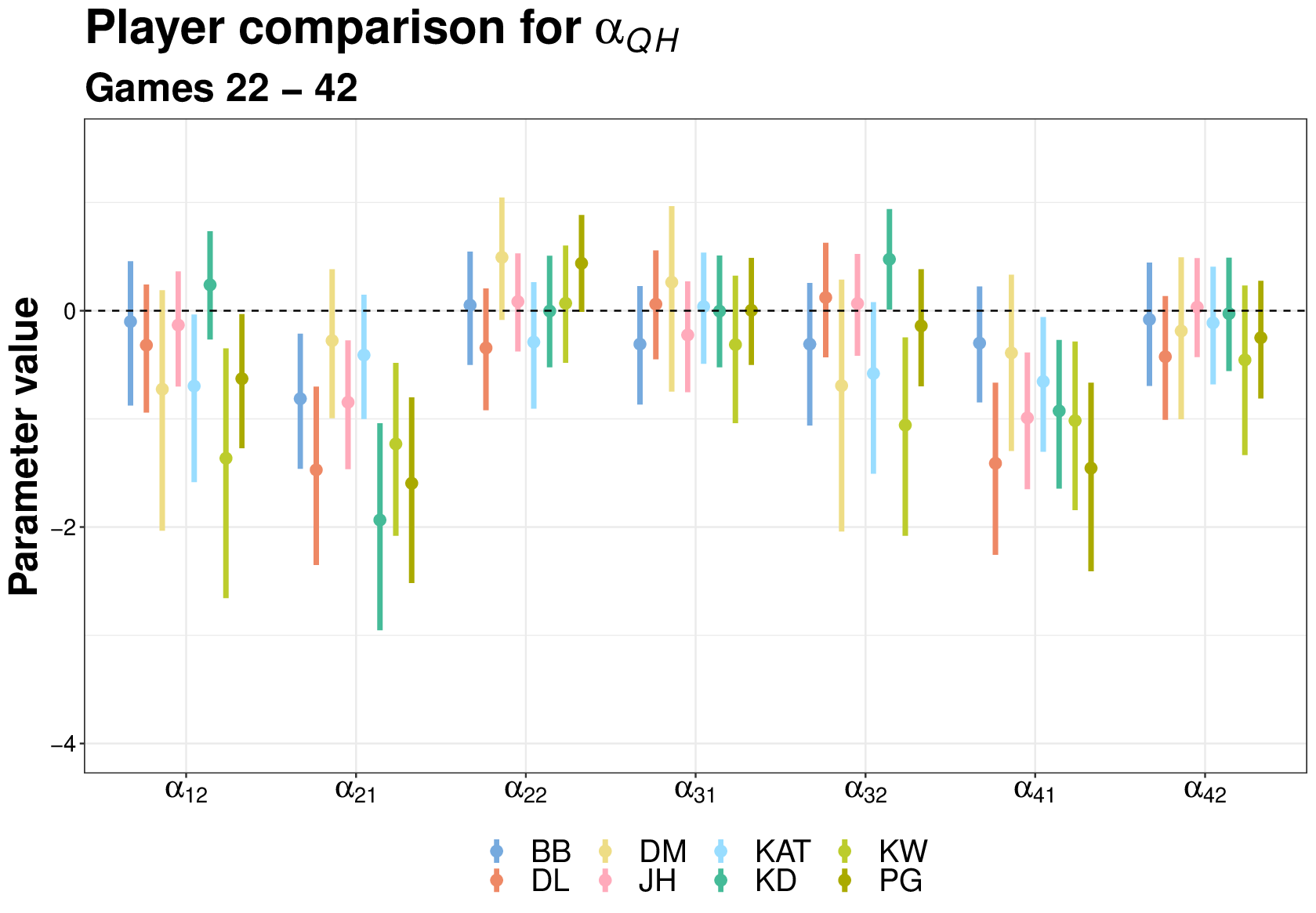}\label{fig:Quarter_half_player_comparison_b}}\\
  \subfloat[]{\includegraphics[width=5cm,angle=-90]{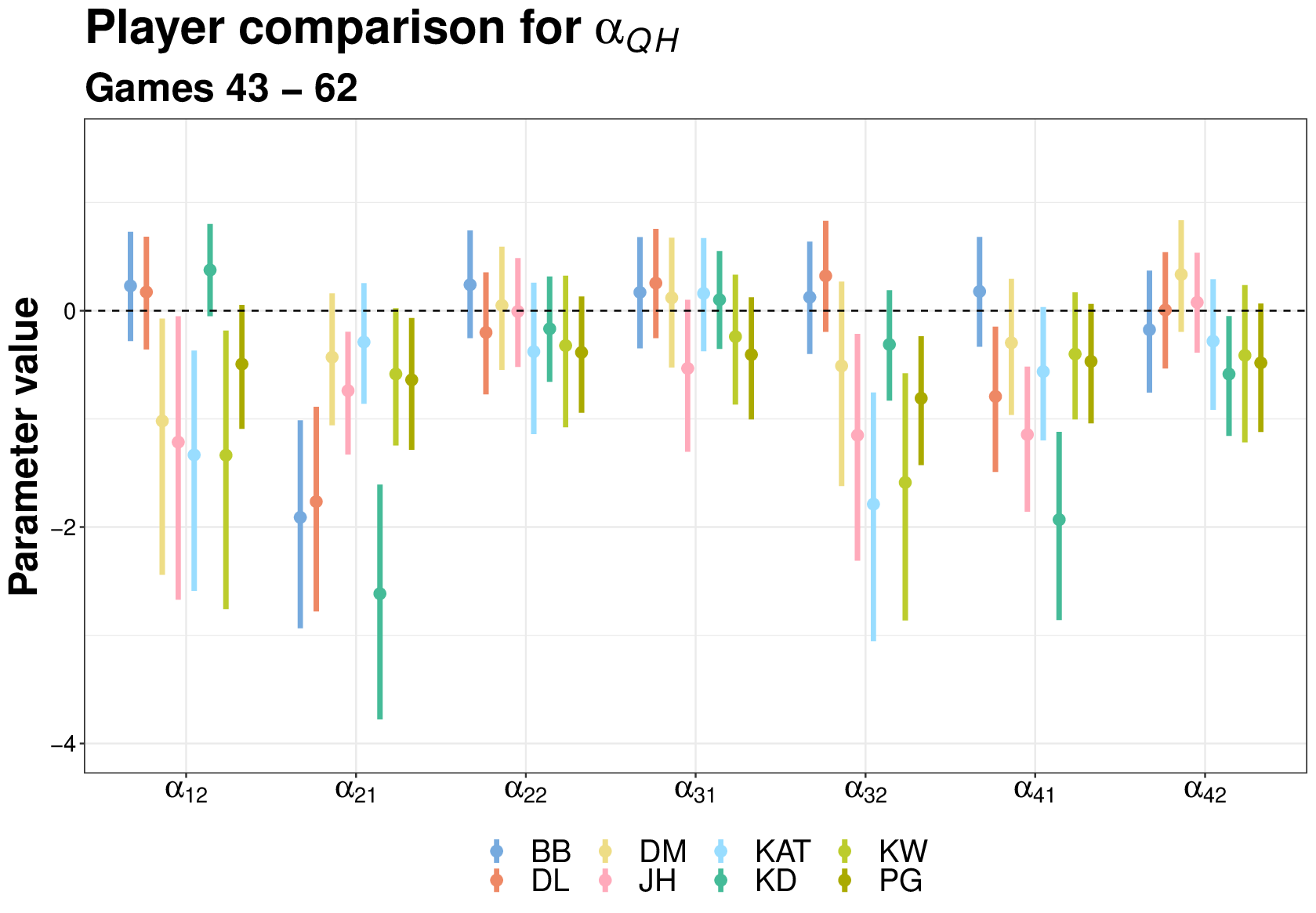}\label{fig:Quarter_half_player_comparison_c}}
 \subfloat[]{\includegraphics[width=5cm,angle=-90]{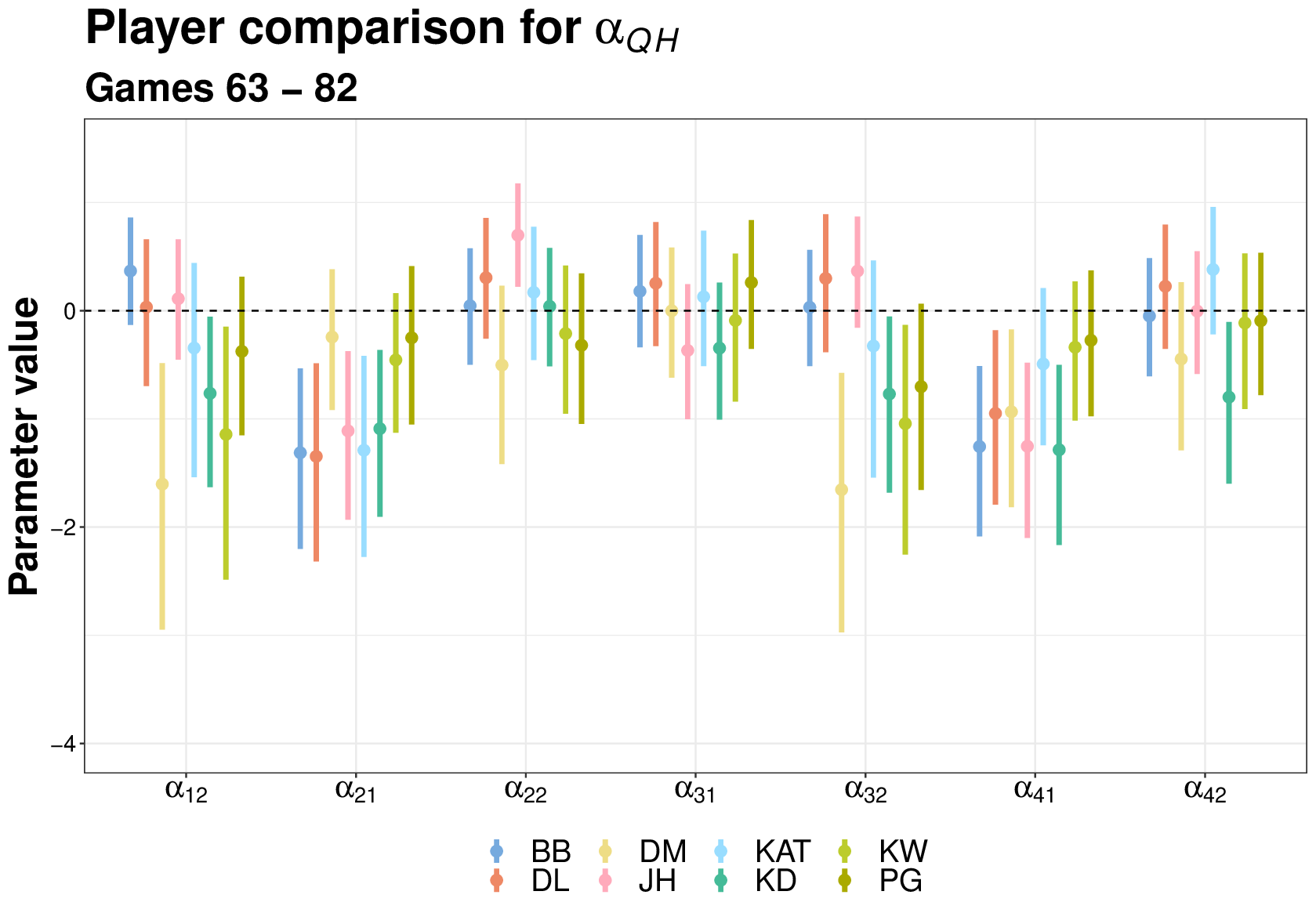}\label{fig:Quarter_half_player_comparison_d}}
\caption{Estimates of the $\alpha_{QH}$'s, computed as the mean value of the posterior distribution of the parameter, and associated 95\% credible intervals for the players under study. The estimates are provided for each of the periods of the season considered for model inference}
\label{fig:Quarter_half_player_comparison}
\end{figure}

The within-game temporal effects are given by the $\alpha_{QH}$ parameters involved in the minute-level modeling approach followed. For this model, we have followed the divide-and-conquer estimation strategy mentioned in previous sections. Particularly, for each team and player under analysis, four minute-level models have been fitted, considering the following four subsets of games of the whole NBA season: $G_1=\{1,...,21\}$, $G_2=\{22,...,42\}$, $G_3=\{43,...,62\}$, and $G_4=\{63,...,N_g\}$. Figures \ref{fig:Quarter_half_team_comparison} and \ref{fig:Quarter_half_player_comparison} show, respectively, the estimates obtained for the $\alpha_{QH}$'s across teams and players for each of these subsets, which allows us to analyze the existence of within-game temporal variability in scoring levels. It is worth noting again that these estimates have to be interpreted in comparison with the first half of the first quarter (reference level), which is the time interval that contains the first six minutes of a game. Most of the estimates of the $\alpha_{QH}$ parameter are negative, which would indicate that the scoring level is lower after minute 6 of the game than before. This fact is fairly consistent among the different teams studied, even though most of the 95\% credible intervals shown in Figure \ref{fig:Quarter_half_team_comparison} contain 0. There are also notable differences between the different subsets of the season under consideration. For example, DEN shows a lower average scoring level in the $Q_{41}$ period only at some points in the season. On the other hand, the players analyzed show lower scoring levels in the $Q_{21}$ and $Q_{41}$ periods, being this result statistically relevant in most cases. This may be due, especially in the case of $Q_{21}$, to the fact that this period of the game is usually a rest period for these players. Estimating the model by segmenting the data set allows, in addition to reducing the computational complexity, to observe the existence of variations in the estimates throughout the season. In order to have a season-level estimate of the $\alpha_{QH}$'s, we could find the barycentric posterior distribution given the four available posterior distributions for each parameter and team/player. This step is going to be exemplified in the next section, considering the self-exciting effects of the models.

\subsection{Self-exciting effects}

This section contains the results obtained for the self-exciting effects of the model at both the within-season and the within-game level. As a summary, Table \ref{etas_kappas_table} shows the posterior mean of the $\eta_S$'s ($\eta_G$'s) and the $\kappa_S$'s ($\kappa_G$'s) for both teams and players, and the corresponding estimates of $\eta_{S}+\kappa_{S}$ ($\eta_{G}+\kappa_{G}$). These are below 1 in all cases, and the estimates of $P(\eta_S+\kappa_S<1)$ ($P(\eta_G+\kappa_G<1)$) derived from MCMC sampled values are equal to 1, confirming the stationarity of the model, which is guaranteed by the condition $\eta_{S}+\kappa_{S}<1$ ($\eta_{G}+\kappa_{G}<1$). 

\begin{table}[hbt]
\centering
\begin{tabular}{lcccccc}
  \cline{2-7}
 & \multicolumn{3}{c}{\textbf{Game-level models}} & \multicolumn{3}{c}{\textbf{Minute-level models}}\\ 
  \hline
\textbf{Team/Player name} & \boldsymbol{$\hat{\eta}_S$} & \boldsymbol{$\hat{\kappa}_S$} & \boldsymbol{$\hat{\eta}_S+\hat{\kappa}_S$} & \boldsymbol{$\hat{\eta}_G$} & \boldsymbol{$\hat{\kappa}_G$} & \boldsymbol{$\hat{\eta}_G+\hat{\kappa}_G$}\\ 
  \hline
Boston Celtics & 0.13 & 0.12 & 0.24 & 0.03 & 0.16 & 0.20 \\ 
Denver Nuggets & 0.11 & 0.18 & 0.28 & 0.03 & 0.30 & 0.33 \\ 
Golden State Warriors & 0.14 & 0.11 & 0.24 & 0.04 & 0.23 & 0.27 \\ 
Houston Rockets & 0.12 & 0.10 & 0.22 & 0.03 & 0.25 & 0.29 \\ 
Milwaukee Bucks & 0.09 & 0.11 & 0.19 & 0.04 & 0.22 & 0.26 \\ 
Philadelphia 76ers & 0.10 & 0.16 & 0.26 & 0.04 & 0.26 & 0.30 \\ 
Portland Trail Blazers & 0.10 & 0.11 & 0.20 & 0.03 & 0.27 & 0.30 \\ 
Toronto Raptors & 0.09 & 0.10 & 0.19 & 0.03 & 0.29 & 0.33 \\ \hline
Bradley Beal & 0.14 & 0.24 & 0.38 & 0.05 & 0.15 & 0.21 \\ 
Damian Lillard & 0.05 & 0.24 & 0.29 & 0.06 & 0.14 & 0.20 \\ 
Donovan Mitchell & 0.10 & 0.27 & 0.37 & 0.06 & 0.51 & 0.57 \\ 
James Harden & 0.26 & 0.27 & 0.54 & 0.05 & 0.17 & 0.21 \\ 
Karl-Anthony Towns & 0.16 & 0.58 & 0.74 & 0.09 & 0.32 & 0.41 \\ 
Kemba Walker & 0.08 & 0.11 & 0.19 & 0.09 & 0.48 & 0.57 \\ 
Kevin Durant & 0.20 & 0.19 & 0.39 & 0.04 & 0.14 & 0.18  \\ 
Paul George & 0.28 & 0.20 & 0.48 & 0.06 & 0.18 & 0.24 \\ 
   \hline
\end{tabular}
\caption{Posterior mean of $\eta_S$, $\kappa_S$, $\eta_G$, and $\kappa_G$ for each of the teams and players under analysis. The condition $\hat{\eta}_{S}+\hat{\kappa}_{S}<1$ ($\hat{\eta}_{G}+\hat{\kappa}_{G}<1$) indicates the stationarity of the game-level (minute-level) model}
\label{etas_kappas_table}
\end{table}

\subsubsection{Within-season self-exciting effects}

\begin{figure}[hbt]
 \centering
 \subfloat[]{\includegraphics[width=5cm,angle=-90]{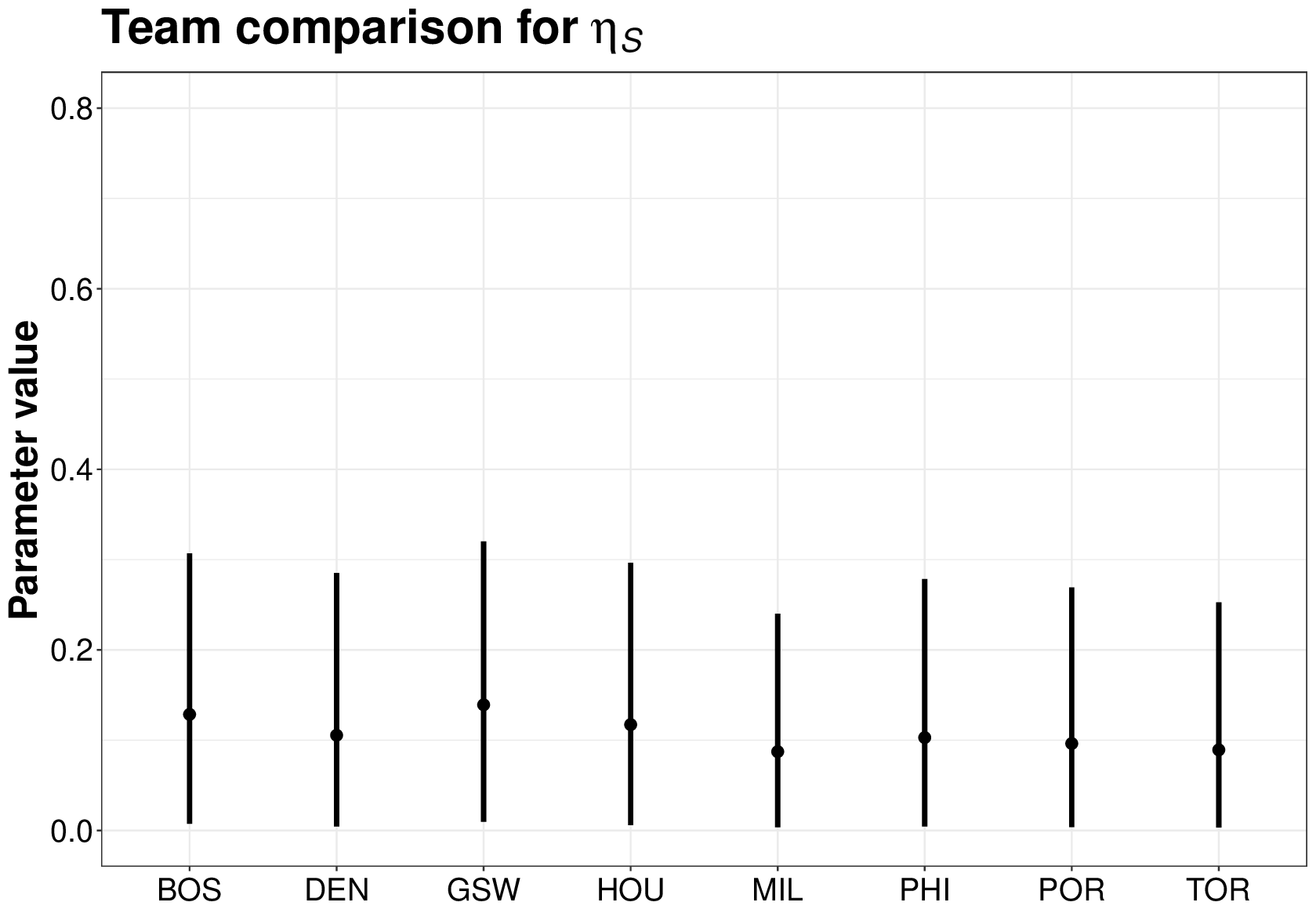}\label{fig:eta_S_kappa_S_teams_a}}
 \subfloat[]{\includegraphics[width=5cm,angle=-90]{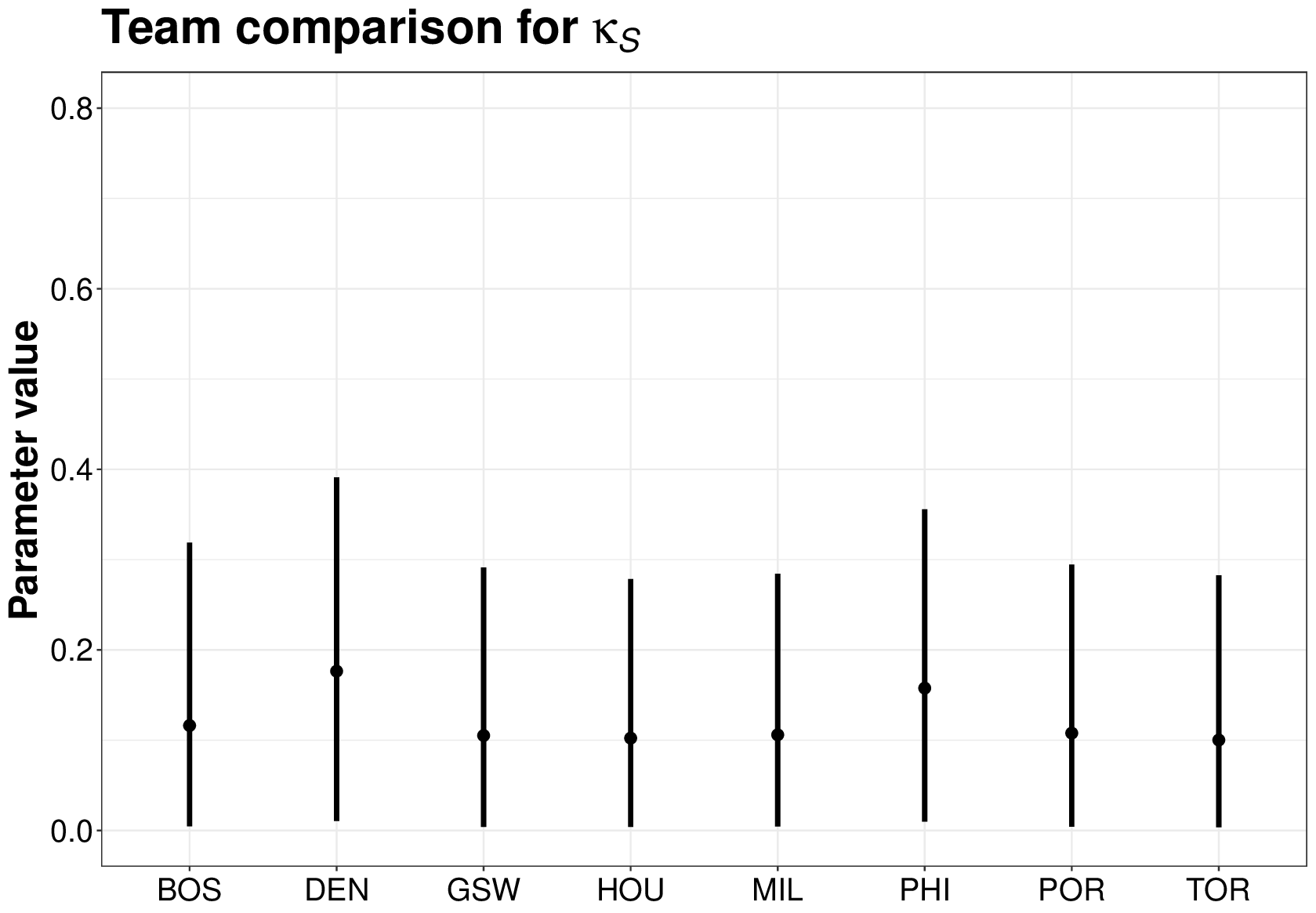}\label{fig:eta_S_kappa_S_teams_b}}\\
\caption{Estimates of the $\eta_S$'s and the $\kappa_S$'s, computed as the mean value of the posterior distribution of the parameter, and associated 95\% credible intervals for the teams under study}
\label{fig:eta_S_kappa_S_teams}
\end{figure}

\begin{figure}[hbt]
 \centering
 \subfloat[]{\includegraphics[width=5cm,angle=-90]{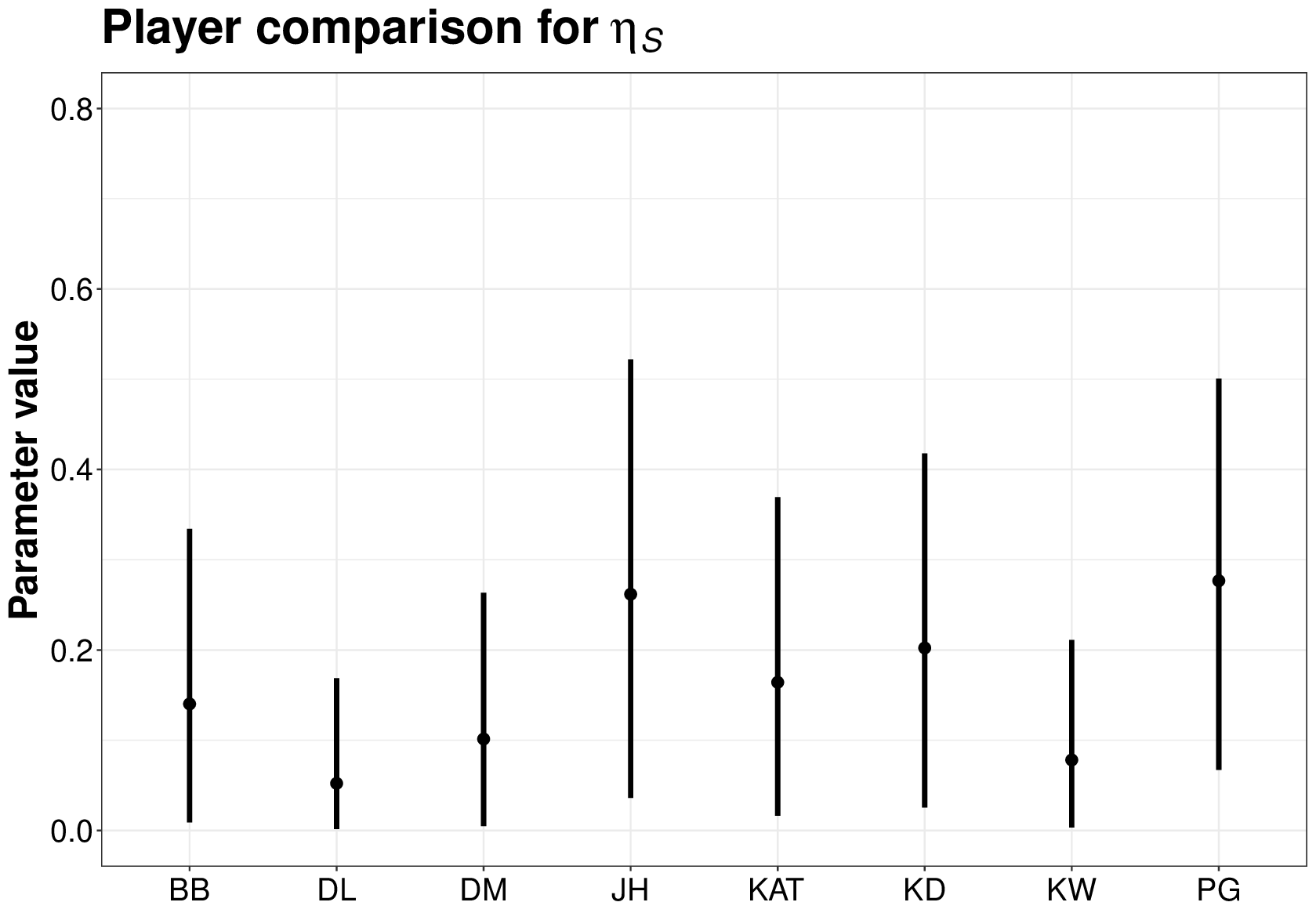}\label{fig:eta_S_kappa_S_players_a}}
 \subfloat[]{\includegraphics[width=5cm,angle=-90]{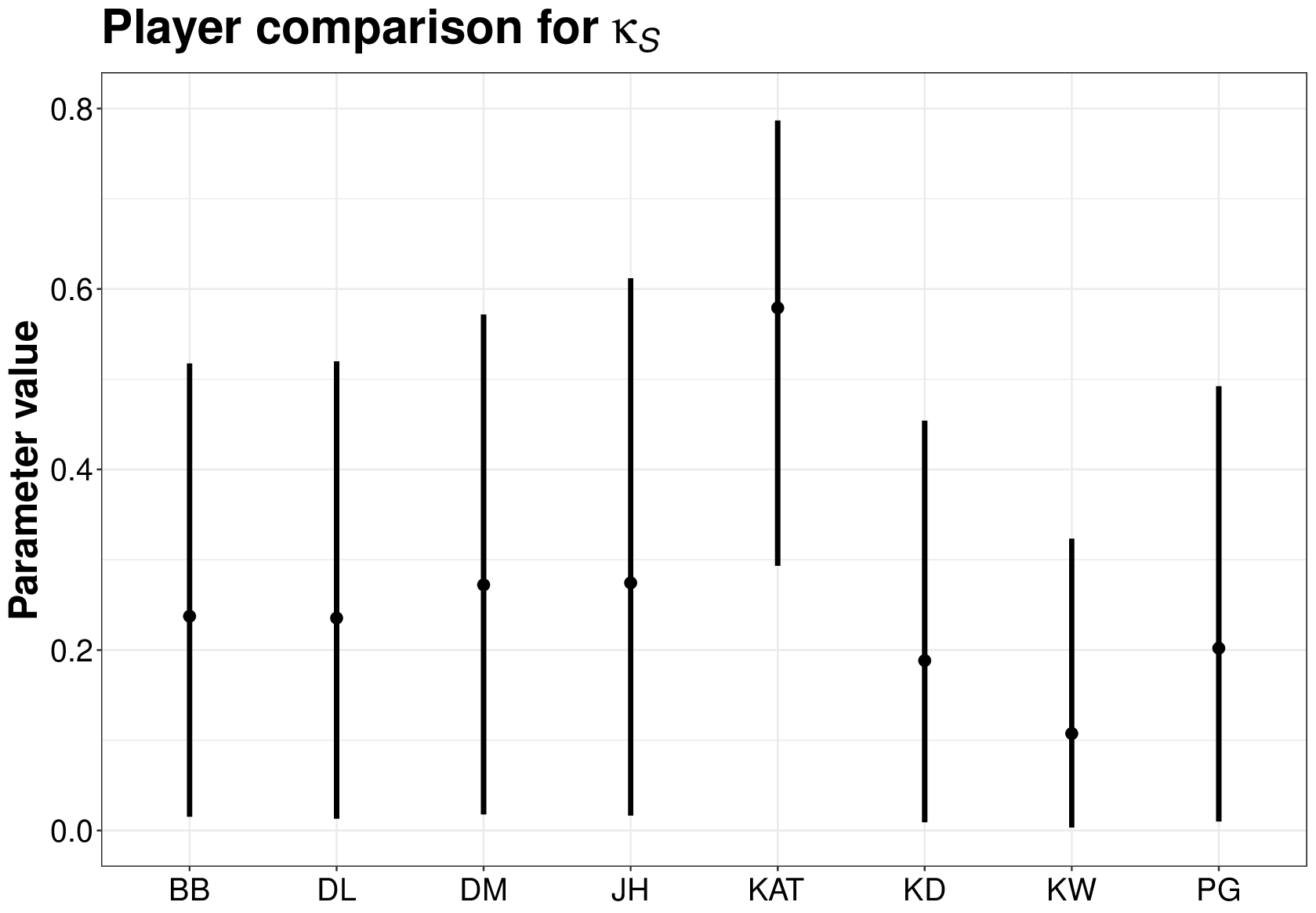}\label{fig:eta_S_kappa_S_players_b}}\\
\caption{Estimates of the $\eta_S$'s and the $\kappa_S$'s, computed as the mean value of the posterior distribution of the corresponding parameter, and associated 95\% credible intervals for the players under study}
\label{fig:eta_S_kappa_S_players}
\end{figure}

\begin{figure}[hbt]
 \centering
\includegraphics[width=8cm,angle=-90]{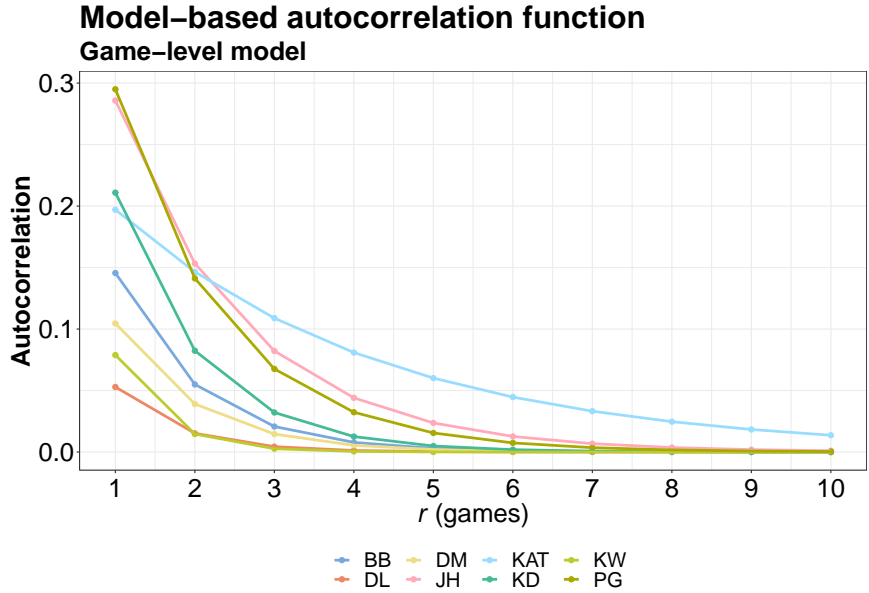}
\caption{Estimation of the autocorrelation function based on the game-level self-exciting model fitted for each player under analysis}
\label{fig:correlation_players}
\end{figure}

Regarding the study of self-exciting effects at the game level, we must focus on the estimation of the parameters $\eta_S$ and $\kappa_S$ of the model. Figures \ref{fig:eta_S_kappa_S_teams} and \ref{fig:eta_S_kappa_S_players} show the posterior means and the associated 95\% credible intervals for both parameters corresponding to the teams (Figure \ref{fig:eta_S_kappa_S_teams}) and players (Figure \ref{fig:eta_S_kappa_S_players}) under analysis. It can be observed that the estimates for both $\eta_S$ and $\kappa_S$ barely differ across teams, even though the higher value of the posterior mean of $\kappa_S$ for DEN and PHI could be highlighted. The situation is quite different among the players, who do show considerably distinct estimates for these two parameters. Indeed, it can be seen that JH and PG present a higher value for $\eta_S$ than the rest, while KAT stands out for its high value of $\kappa_S$. At the same time, some players such as DL or KW present small estimates for $\eta_S$, and also for $\kappa_S$ in the case of KW. In order to visualize how the estimates of $\eta_S$ and $\kappa_S$ reflect different scoring tendencies for these players, Figure \ref{fig:correlation_players} shows the estimation of the autocorrelation function that the fitted models yield (according to (\ref{eq:autocorrelation_function})), given the estimates of $\eta_S$ and $\kappa_S$. This autocorrelation function is estimated for lag values $r=1,...,10$, representing the correlation coefficient between $Y_g$ and $Y_{g-r}$. In particular, Figure \ref{fig:correlation_players} enables us to observe that JH and PG present an estimated value of the autocorrelation function close to 0.3 for $r=1$, which is the highest among the players analyzed, corresponding to the greater values of $\eta_S$ estimated for these two players. We can also appreciate that the temporal dependence in game-level scoring levels remains longer for KAT, which is the player that presents the highest value for $\kappa_S$. In any case, we should note that the credibility intervals associated with these estimates overlap, so the differences found between players should be taken with caution. Nevertheless, the analysis of the model at the game level allows us to conclude that although game-level scoring dynamics do not characterize very well the teams under study, some of the best scorers in the league show noticeable differences in this regard.

\subsubsection{Within-game self-exciting effects}

\begin{figure}[hbt]
\centering
\includegraphics[width=8cm,angle=-90]{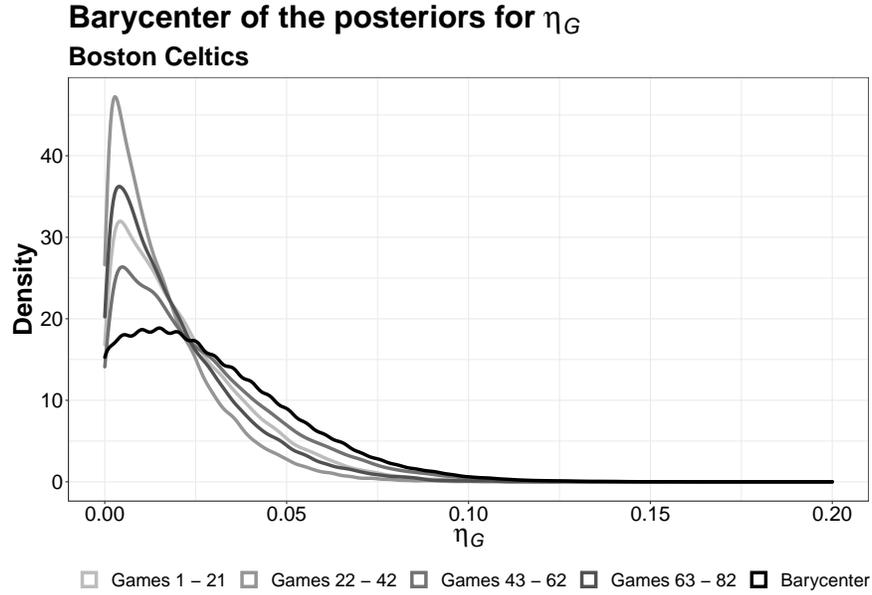}
\caption{Posterior distribution of $\eta_G$ corresponding to the Boston Celtics and each of the periods of the season considered for the analysis, and barycenter of these posterior distributions based on the Wasserstein distance}
\label{fig:barycenter}
\end{figure}

The study of within-game self-exciting effects is based on the $\eta_G$ and the $\kappa_G$ parameters estimated through the minute-level model. We compute the barycentric posterior distribution of these parameters considering the four subsets of the NBA season previously mentioned, which results in a unique season-level estimate of these parameters. For instance, Figure \ref{fig:barycenter} shows the posterior distribution of $\eta_G$ corresponding to the Boston Celtics for each subset of the season and the barycentric posterior distribution that results in this case. 

\begin{figure}[hbt]
 \centering
 \subfloat[]{\includegraphics[width=5cm,angle=-90]{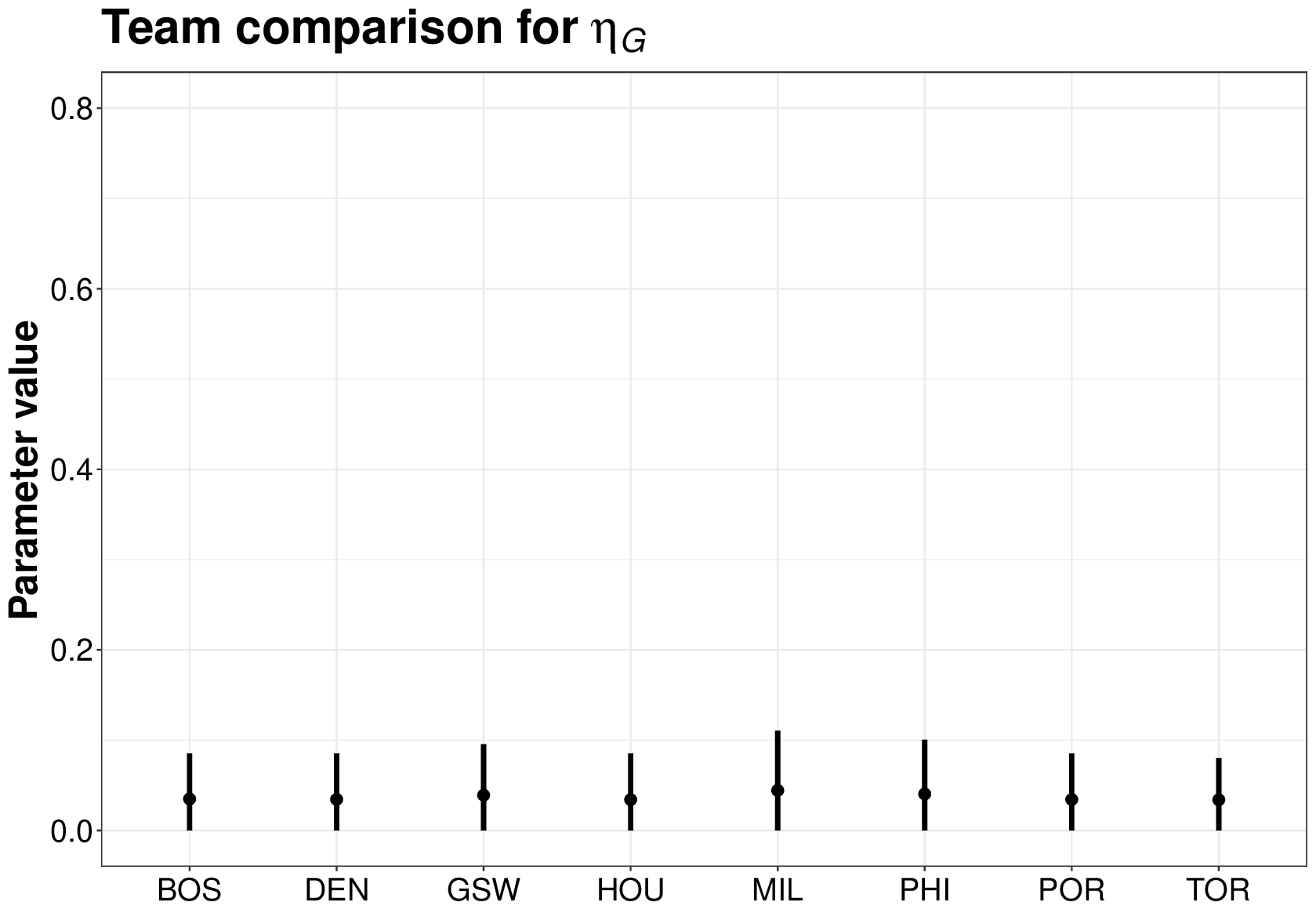}\label{fig:eta_G_kappa_G_teams_a}}
 \subfloat[]{\includegraphics[width=5cm,angle=-90]{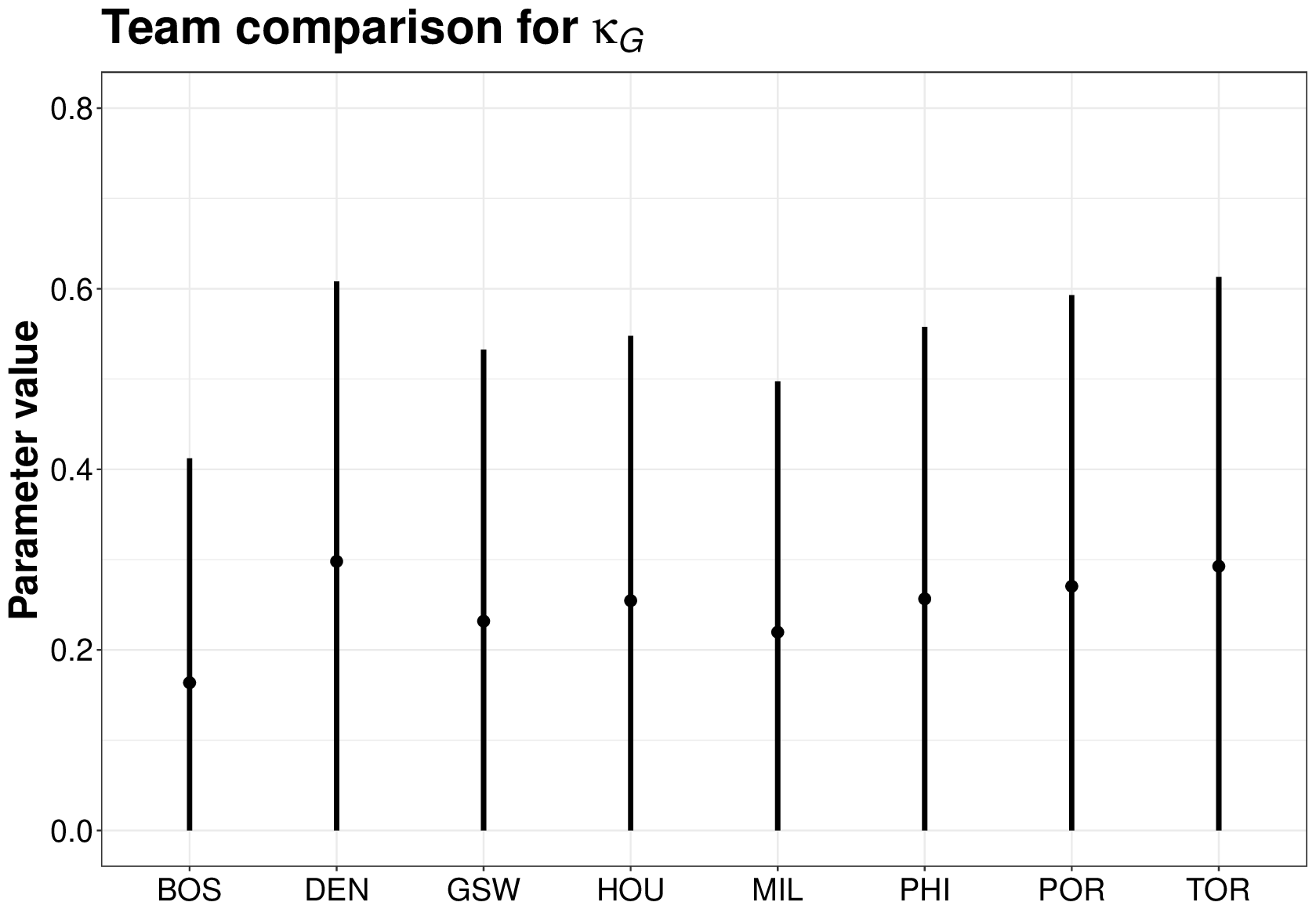}\label{fig:eta_G_kappa_G_teams_b}}\\
\caption{Estimates of the $\eta_G$'s (a) and the $\kappa_G$'s (b), computed as the mean value of the barycentric posterior distribution of the corresponding parameter, and associated 95\% credible intervals for the teams under study}
\label{fig:eta_G_kappa_G_teams}
\end{figure}

\begin{figure}[hbt]
 \centering
 \subfloat[]{\includegraphics[width=5cm,angle=-90]{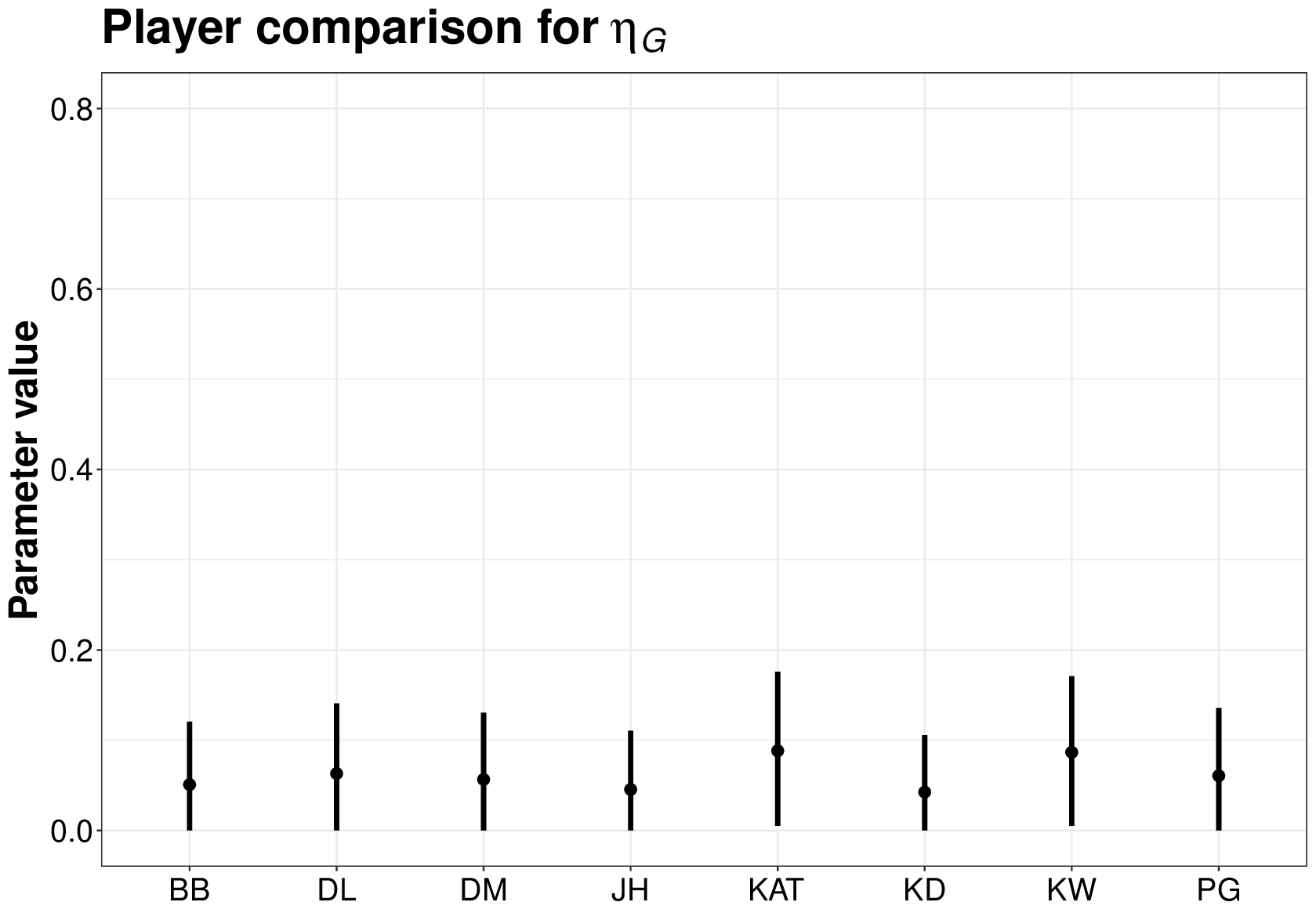}\label{fig:eta_G_kappa_G_players_a}}
 \subfloat[]{\includegraphics[width=5cm,angle=-90]{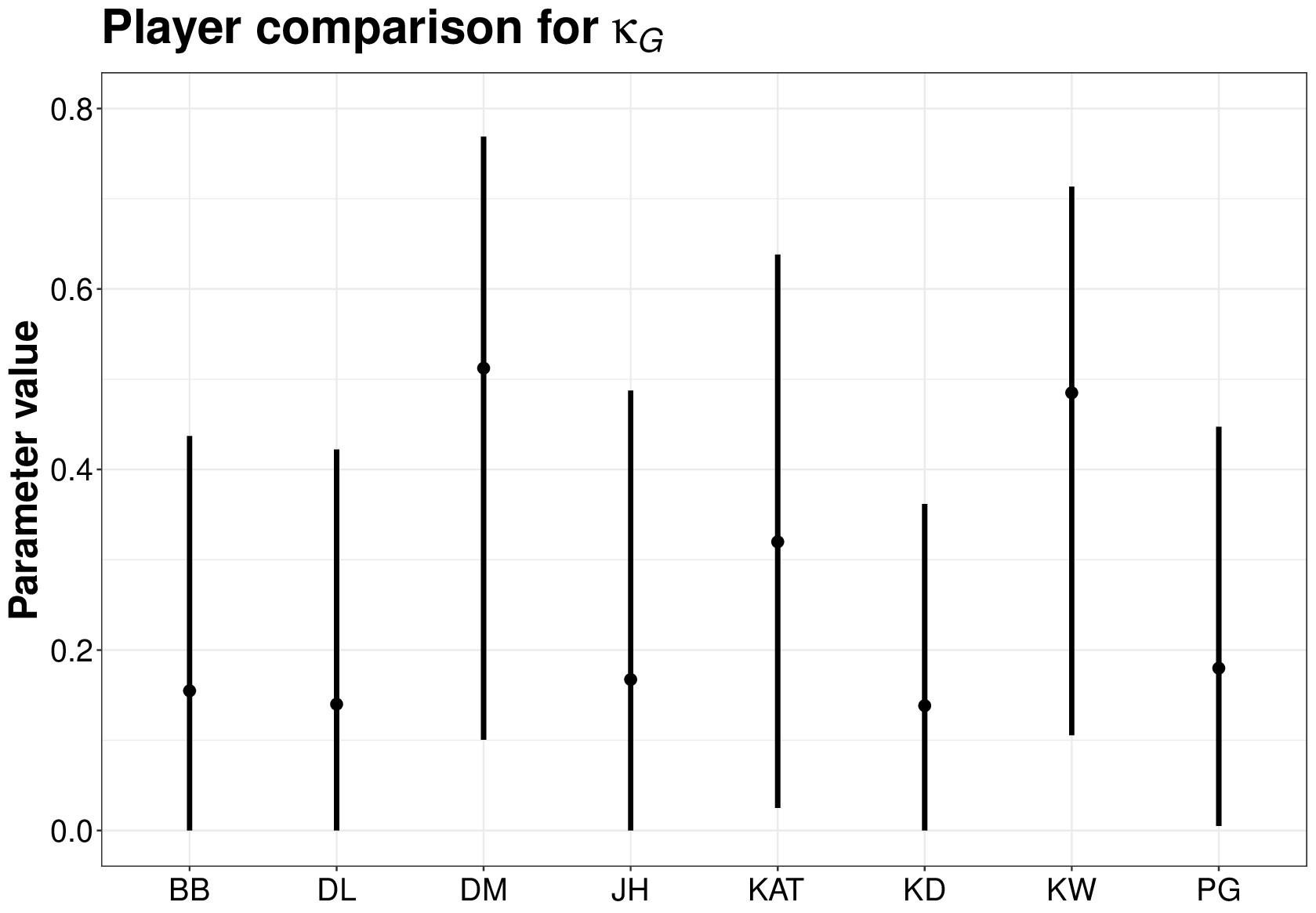}\label{fig:eta_G_kappa_G_players_b}}\\
\caption{Estimates of the $\eta_G$'s (a) and the $\kappa_G$'s, computed as the mean value of the barycentric posterior distribution of the corresponding parameter, and associated 95\% credible intervals for the players under study}
\label{fig:eta_G_kappa_G_players}
\end{figure}

\begin{figure}[hbt]
 \centering
\includegraphics[width=8cm,angle=-90]{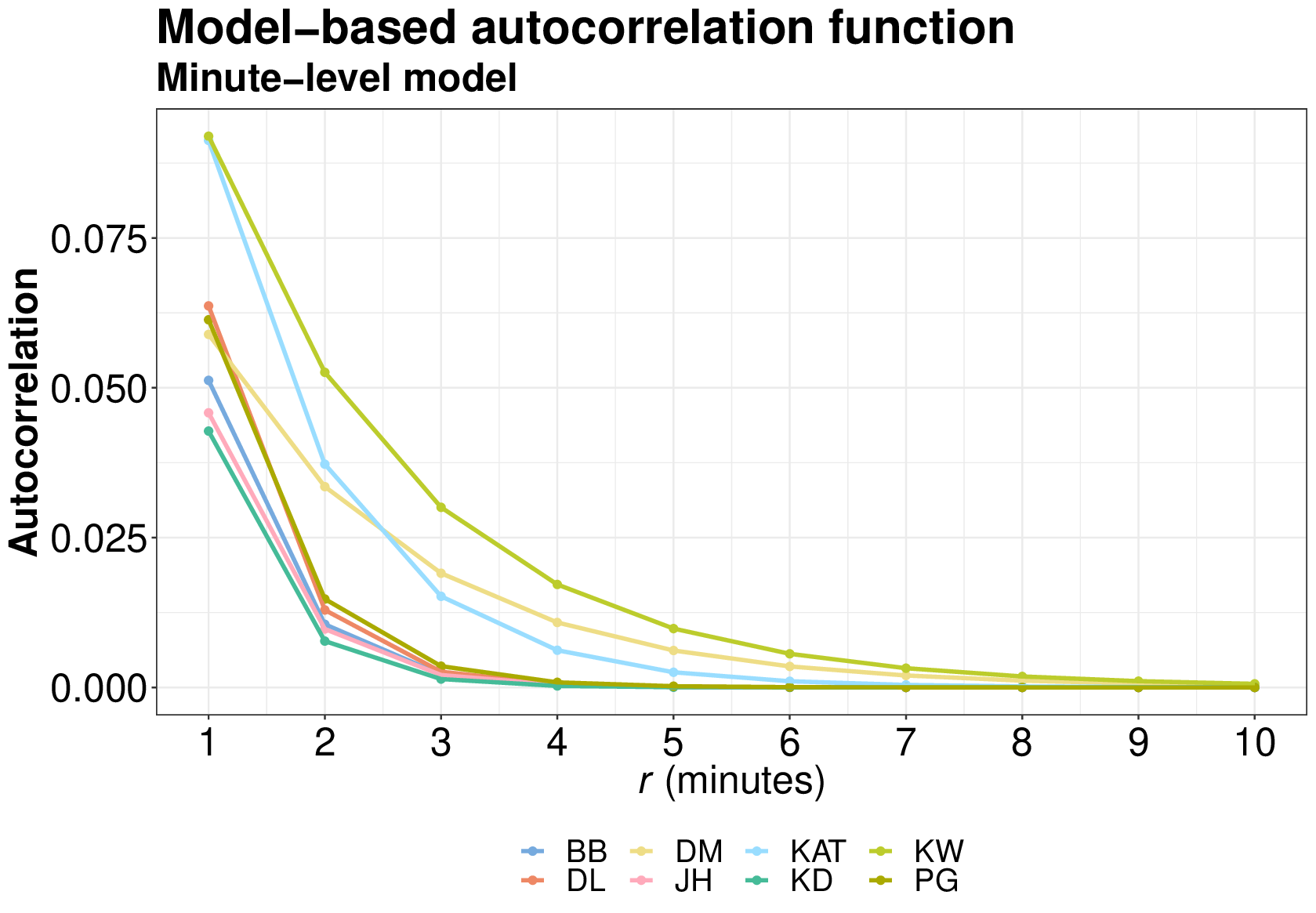}
\caption{Estimation of the autocorrelation function based on the minute-level self-exciting model fitted for each player under analysis}
\label{fig:correlation_players_minute}
\end{figure}

Hence, Figures \ref{fig:eta_G_kappa_G_teams} and \ref{fig:eta_G_kappa_G_players} display the barycentric posterior means and the associated 95\% credible intervals of both parameters for the teams (Figure \ref{fig:eta_G_kappa_G_teams}) and players (Figure \ref{fig:eta_G_kappa_G_players}) under analysis. As in the case of the game-level model, the variability of the $\eta_G$'s and the $\kappa_G$'s is lower across teams rather than players. In view of the season-level estimates obtained from the barycentric posterior distributions, we can highlight that KAT and KW present the greatest values for $\eta_G$, whereas DM, KAT, and KW display the greatest estimates for $\kappa_G$. This result suggests that these three players present a more singular scoring pattern in terms of repetition. Figure \ref{fig:correlation_players_minute} displays the autocorrelation function estimated for the minute-level model (following (\ref{eq:autocorrelation_function})), considering lag values $r=1,...,10$ (in minutes), representing the correlation coefficient between $Y_{gm}$ and $Y_{gm-r}$. It is clear that the autocorrelation function at the minute-level presents lower values than the game-level counterpart. Nevertheless, it helps to understand the magnitude of the temporal dependence between the scoring levels per minute and the existing variations across players.

\subsection{The proposed model vs. a baseline model}

Table \ref{WAIC} shows the WAIC values for each of the game-level and minute-level models fitted, considering the doubly self-exciting specification proposed in the paper, along with the WAIC of the baseline models considered for comparison purposes. These baseline models exclude all the parameters involved in the self-exciting component of the models suggested. Specifically, the game-level baseline model only accounts for the playing-at-home effect, whereas the minute-level baseline model only includes the within-game temporal effects given by the $\alpha_{QH}$'s. Besides, the posterior mean of $\lambda_g$ provided by the game-level baseline model is included as the offset term of the minute-level baseline model. First, the results for the game-level models indicate that considering a self-exciting structure for modeling game-level scoring levels is not beneficial in most cases. In particular, there is no improvement in terms of the WAIC, except for three players: JH, KAT, and PG. This result is consistent with the fact that these players have presented higher estimates for the $\eta_S$ and $\kappa_S$ parameters. Second, the values of the WAIC obtained for the minute-level models suggest that this modeling approach can provide better results, both at the team and player level, in general. Indeed, the doubly self-exciting model provides an improvement in the WAIC for many players/teams and periods within the season. Particularly, for some of them, there is an improvement in at least two periods of the season. In the case of KAT, the improvement is observed in all periods except for the last one. This again suggests the presence of a higher level of self-excitation in this player's scoring levels.

\begin{table}[hbt]
\resizebox{\textwidth}{!}{
\centering
\begin{tabular}{lcccccccccc}
  \cline{2-11}
 & \multicolumn{2}{c}{\multirow{2}{*}{\textbf{Game-level models}}} & \multicolumn{8}{c}{\textbf{Minute-level models}} \\
 &  &  & \multicolumn{2}{c}{\textbf{Games 1-21}} & \multicolumn{2}{c}{\textbf{Games 22-42}}
    & \multicolumn{2}{c}{\textbf{Games 43-62}} & \multicolumn{2}{c}{\textbf{Games 63-82}}\\
  \hline
\textbf{Team/Player name} & \textbf{Baseline} & \textbf{DSE} & \textbf{Baseline} & \textbf{DSE} & \textbf{Baseline} & \textbf{DSE} & \textbf{Baseline} & \textbf{DSE} & \textbf{Baseline} & \textbf{DSE}\\ 
  \hline
Boston Celtics & 499.56 & 504.30 & 2258.38 & 2261.91 & \textbf{2293.57} & \textbf{2292.99} & \textbf{2197.91} & \textbf{2197.75} & 2210.12 & 2210.74 \\
Denver Nuggets & 502.74 & 504.78 & \textbf{2275.48} & \textbf{2274.34} & 2262.05 & 2262.48 & 2213.70 & 2214.11 & 2160.58 & 2161.39 \\
Golden State Warriors & 512.05 & 515.93 & 2303.43 & 2304.84 & \textbf{2305.82} & \textbf{2305.29} & 2251.63 & 2252.79 & \textbf{2201.46} & \textbf{2200.68} \\
Houston Rockets & 497.80 & 501.79 & 2227.13 & 2228.85 & 2205.88 & 2206.22 & 2141.46 & 2141.75 & \textbf{2158.17} & \textbf{2157.49} \\ 
Milwaukee Bucks & 507.45 & 511.67 & 2362.24 & 2364.40 & \textbf{2275.33} & \textbf{2275.03} & \textbf{2218.26} & \textbf{2217.96} & 2205.60 & 2205.77 \\ 
Philadelphia 76ers & 507.70 & 509.82 & 2209.47 & 2210.44 & 2304.81 & 2306.04 & 2218.63 & 2218.92 & \textbf{2160.79} & \textbf{2160.62} \\ 
Portland Trail Blazers & 498.68 & 503.48 & 2260.74 & 2264.26 & \textbf{2258.70} & \textbf{2258.57} & \textbf{2200.71} & \textbf{2200.42} & 2223.38 & 2223.71 \\ 
Toronto Raptors & 501.63 & 507.02 & 2318.89 & 2321.09 & \textbf{2251.91} & \textbf{2251.21} & 2193.42 & 2193.75 & 2183.80 & 2184.56 \\ \hline
Bradley Beal & 407.88 & 410.32 & 937.75 & 939.77 & 1106.79 & 1107.53 & 1041.53 & 1041.79 & 939.01 & 939.90 \\ 
Damian Lillard & 395.01 & 399.41 & 954.11 & 955.56 & 922.74 & 922.78 & 933.75 & 933.88 & 815.00 & 815.57 \\ 
Donovan Mitchell & 400.25 & 404.12 & 957.35 & 960.22 & \textbf{945.64} & \textbf{944.44} & \textbf{1020.03} & \textbf{1019.94} & 712.67 & 713.10 \\ 
James Harden & \textbf{391.25} & \textbf{388.63} & \textbf{1023.76} & \textbf{1023.11} & 1173.90 & 1175.32 & 1147.19 & 1148.66 & \textbf{838.13} & \textbf{837.73} \\ 
Karl-Anthony Towns & \textbf{394.67} & \textbf{382.11} & \textbf{893.64} & \textbf{885.82} & \textbf{1019.17} & \textbf{1018.14} & \textbf{1031.85} & \textbf{1031.55} & 721.25 & 721.91 \\ 
Kemba Walker & 452.34 & 457.64 & 1005.01 & 1008.07 & \textbf{927.24} & \textbf{927.03} & \textbf{988.95} & \textbf{988.39} & 986.18 & 986.55 \\ 
Kevin Durant & 383.84 & 386.39 & 1081.21 & 1082.63 & \textbf{1005.51} & \textbf{1003.00} & 939.18 & 939.73 & 682.43 & 684.19 \\ 
Paul George & \textbf{377.04} & \textbf{376.63} & 922.72 & 924.32 & 1044.72 & 1044.80 & \textbf{1028.94} & \textbf{1026.65} & \textbf{729.18} & \textbf{728.60} \\ 
   \hline
\end{tabular}}
\caption{WAIC values obtained for the doubly self-exciting (DSE) model proposed in the paper and a baseline model excluding game-level and minute-level self-exciting effects. The values in bold correspond to those teams/players for which the doubly self-exciting model outperforms the baseline one}
\label{WAIC}
\end{table}

\subsection{Clustering scoring levels}

\begin{figure}[hbt]
 \centering
 \subfloat[]{\includegraphics[width=5cm,angle=-90]{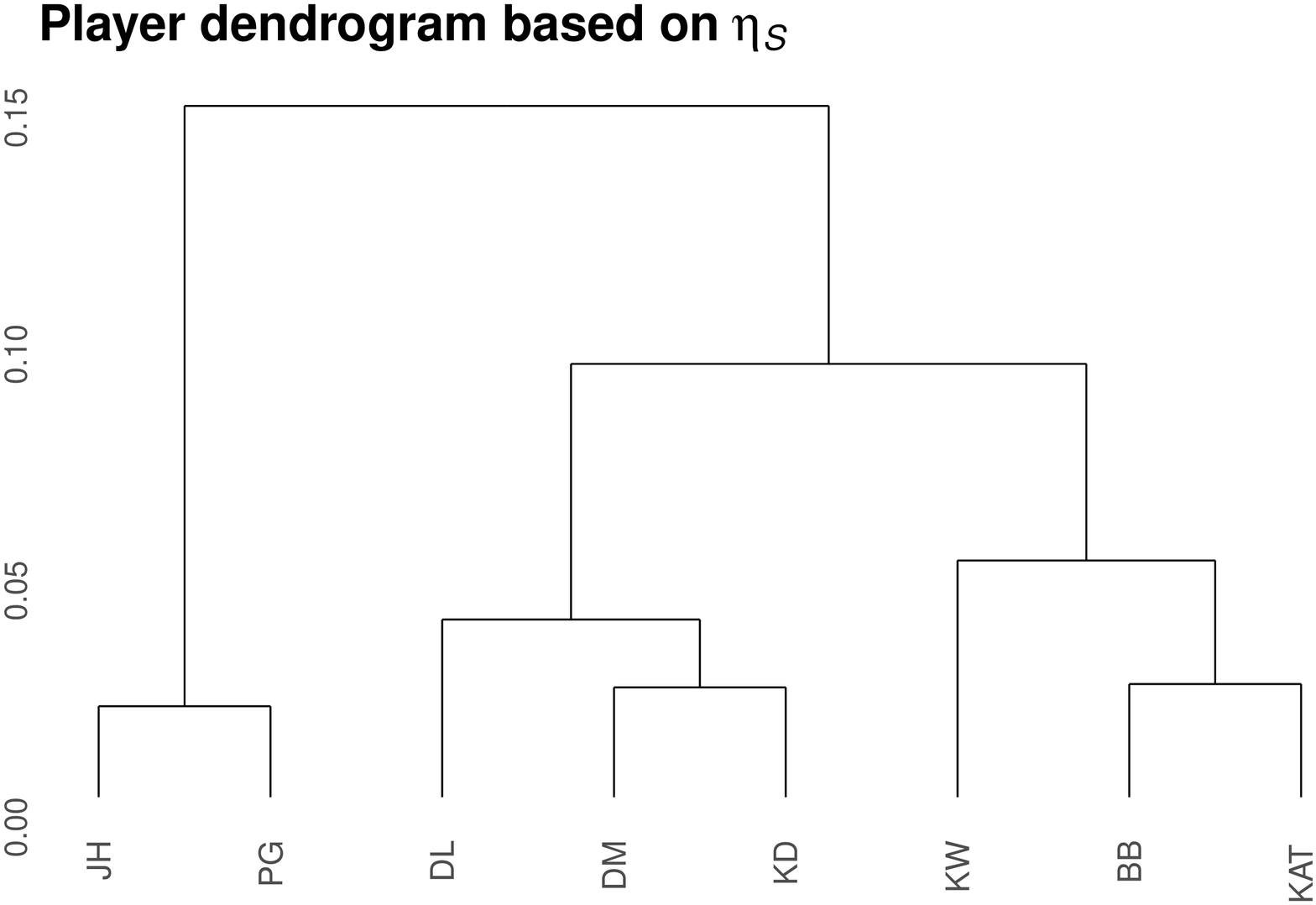}\label{fig:dendros_players_game_model_a}}
 \subfloat[]{\includegraphics[width=5cm,angle=-90]{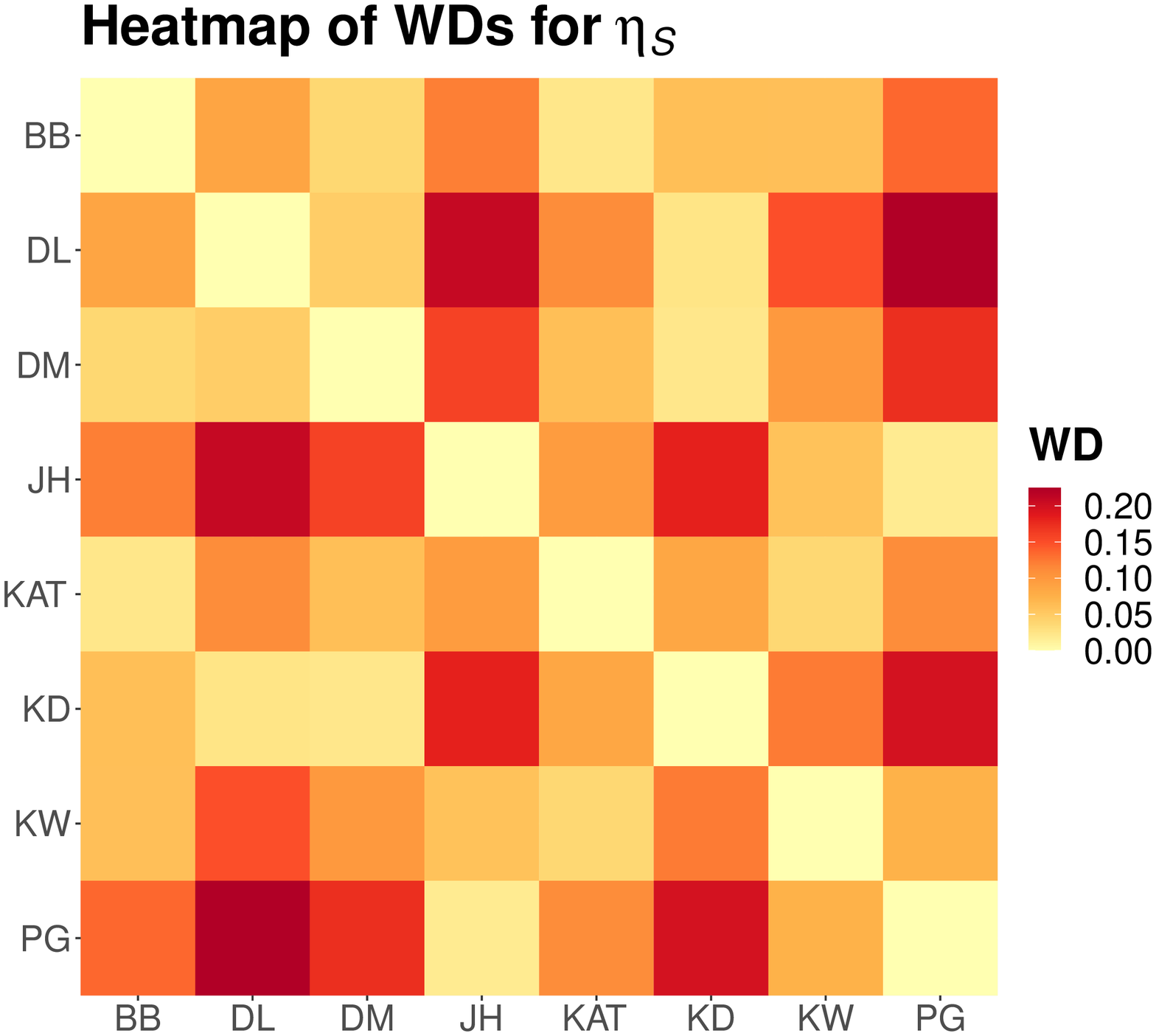}\label{fig:dendros_players_game_model_b}}\\
  \subfloat[]{\includegraphics[width=5cm,angle=-90]{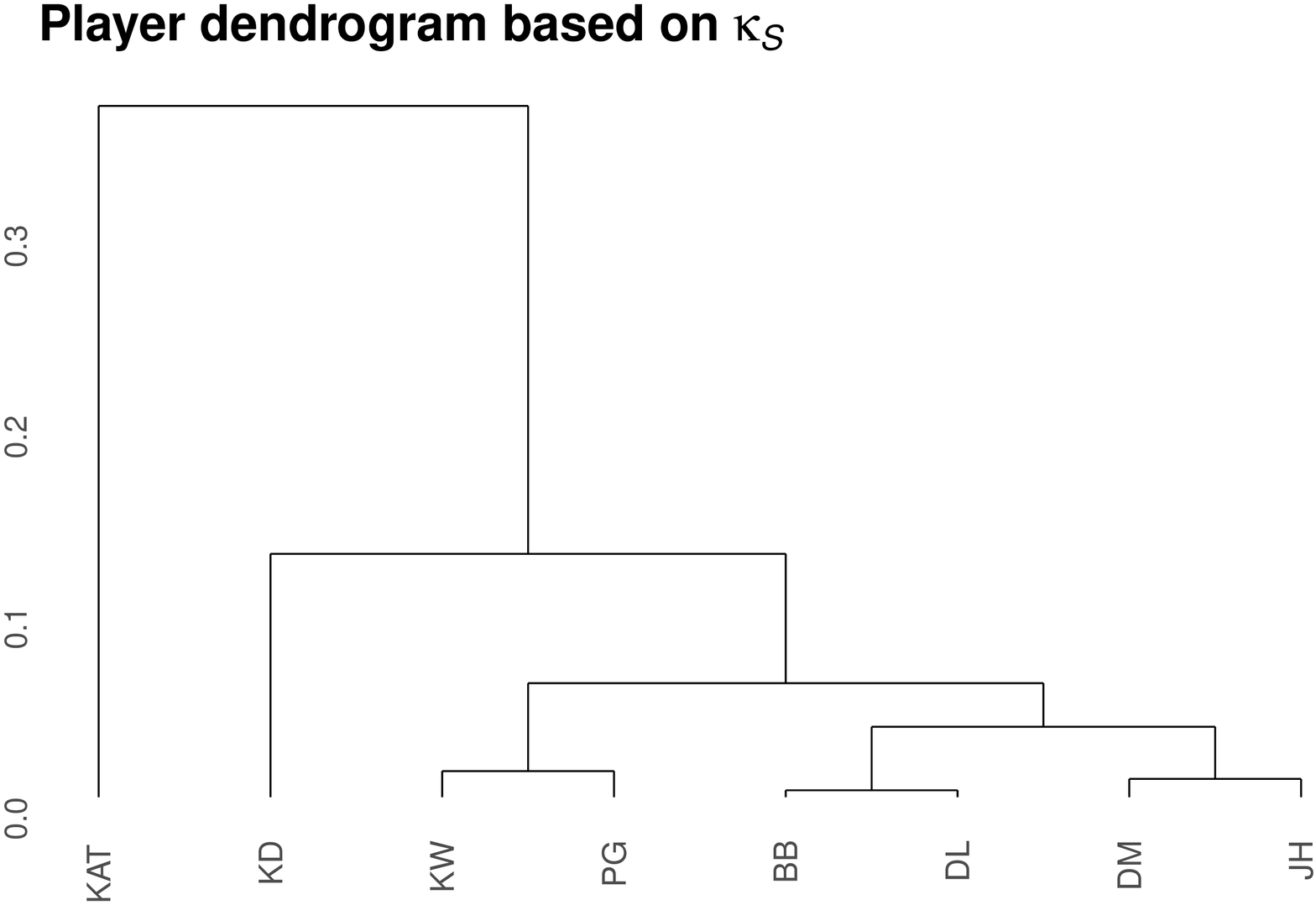}\label{fig:dendros_players_game_model_c}}
 \subfloat[]{\includegraphics[width=5cm,angle=-90]{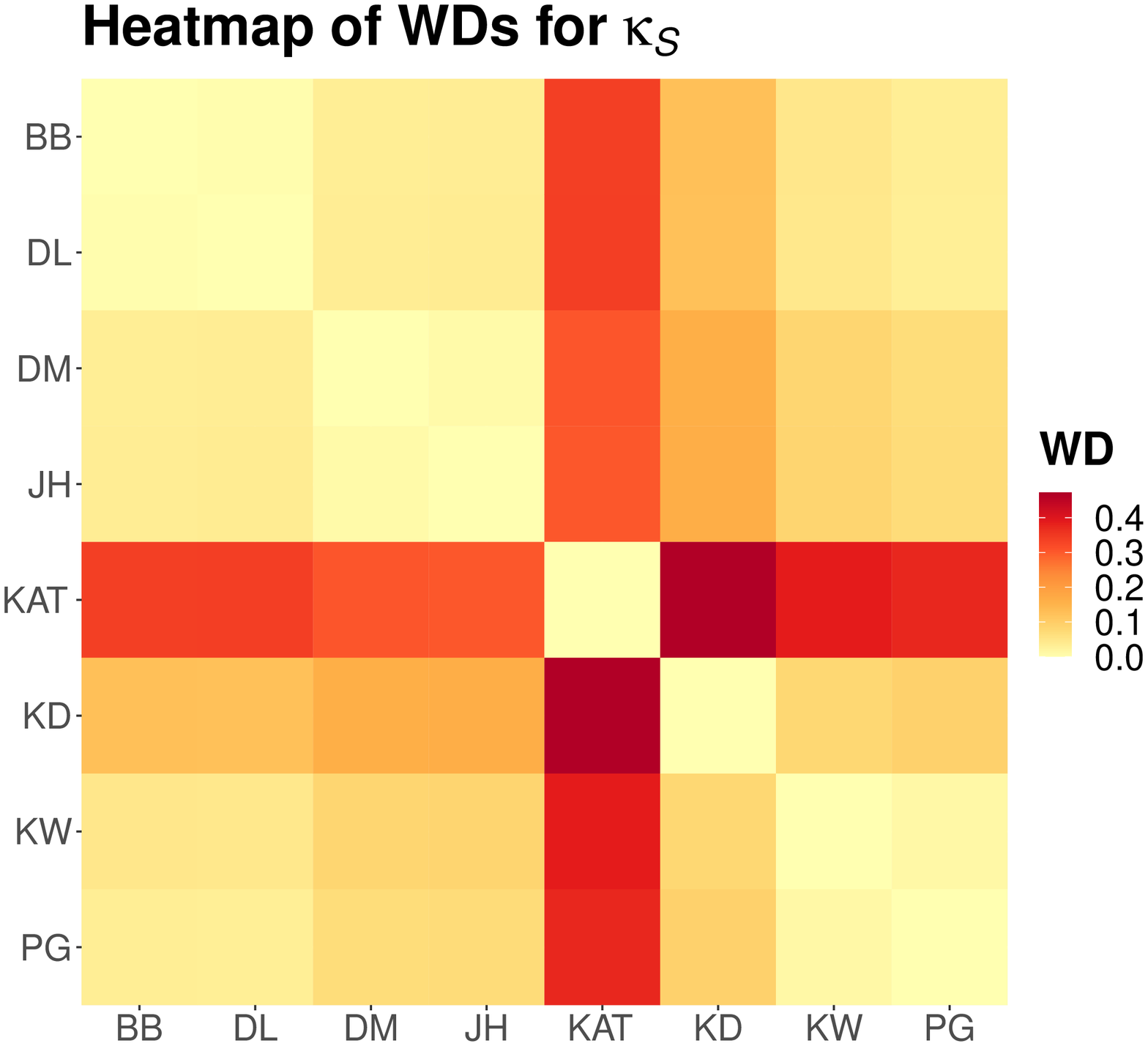}\label{fig:dendros_players_game_model_d}}
\caption{Player dendrogram based on the $\eta_S$ (a) and $\kappa_S$ (c) parameters, along with the dissimilarity matrices (expressed as heatmaps) between players in terms of the Wasserstein distances (WDs) for $\eta_S$ (b) and $\kappa_S$ (d)}
\label{fig:dendros_players_game_model}
\end{figure}

Finally, a hierarchical clustering of the players has been performed based on the estimates obtained for the parameters involved in the self-exciting component of the models. Figure \ref{fig:dendros_players_game_model} shows the corresponding dendrograms for the set of players taking into account the barycentric estimates of the posterior distribution of $\eta_S$ and $\kappa_S$. The dissimilarity matrices resulting from calculating the Wasserstein distance between the posterior distributions of each pair of players are also included. We have chosen $\eta_S$ and $\kappa_S$ for illustrative purposes since they are the self-exciting parameters that show the greatest variability between players, rather than those that correspond to the minute-level model. For the same reason, this analysis is carried out between players and not between teams, which present few differences in terms of $\eta_S$ and $\kappa_S$. In the case of $\eta_S$, JH and PG, the players who show the highest values for this parameter, are also those with the greatest similarity in terms of the corresponding posterior distributions. However, the differences between players are not so large for parameter $\eta_S$. For $\kappa_S$, in contrast, KAT shows a posterior distribution far away from that of the rest, as reflected in the dendrogram (Figure \ref{fig:dendros_players_game_model_c}) and in the heatmap associated to the dissimilarity matrix (Figure \ref{fig:dendros_players_game_model_d}).

\section{Discussion}
\label{Discussion}

In this paper, a doubly self-exciting Poisson model is proposed for the analysis of time series data representing the number of field goals made by a basketball team/player, considering two different time domains. In this sense, the presence of the modifiable temporal unit problem \citep{cheng2014modifiable} cannot be discarded, since the choice of the game and the minute as temporal units of reference, while natural, is arbitrary. Choosing a larger within-game temporal unit could turn out to be convenient.

The existence of contagiousness in scoring levels is an issue of great interest that is often approached under the hot hand hypothesis, which focuses on the percentage of success in field goal or free throw shooting. In this paper, however, we have put the focus on scoring levels in general, meaning somehow the scoring productivity of a team or player. Studying the self-exciting nature of this variable is an alternative approach that also reflects the scoring dynamics that a team or player may experience within the season or a game. The results do not indicate the existence of such self-exciting dynamics for some players or teams, but the model proves to be suitable for others. In particular, the modeling approach followed has shown to be useful for detecting different scoring patterns among the players considered. Clustering teams or players according to the Wasserstein distances between the posterior distributions associated with key parameters has proven to be an inexpensive option in this regard.

Future studies in the same direction could be helpful to unveil within-team team scoring patterns. In particular, the analysis of the multivariate time series corresponding to the scoring patterns of all the players of a team could reveal the existence of within-team contagious dynamics among some players, an idea that could be envisaged under the concept of ``team chemistry'' recently studied by \cite{horrace2022network}. 

Finally, it is worth noting that the model proposed has only been applied for explanatory purposes, to try to describe and understand the scoring patterns of some NBA teams and players. The use of this model as a forecasting in-game tool might also deserve further consideration in the context of basketball or other sports where its application would also be reasonable. 

\clearpage



\section*{\textbf{Data availability statement}}

The NBA dataset used for the analysis and the main R codes used for model fitting will be provided in \url{https://github.com/albrizre/NBA_DSE}.

\section*{\textbf{Funding}}

This research did not receive any specific grant from funding agencies in the public, commercial, or not-for-profit sectors.

\section*{\textbf{Conflict of interest}}

The author declares no conflicts of interest.


\bibliography{Bibliography}

\end{document}